# The domestic economic impacts of immigration

David Roodman[1]

September 2014

**Summary:** This paper critically reviews the research on the impact of immigration on employment and wages of natives in wealthy countries—where "natives" includes previous immigrants and their descendants. While written for a non-technical audience, the paper engages with technical issues and probes one particularly intense scholarly debate in an appendix. While the available evidence is not definitive, it paints a consistent picture. Industrial economies can generally absorb migrants quickly, in part because capital is mobile and flexible, in part because immigrants are consumers as well as producers. Thus, long-term average impacts are probably zero at worst. And the long-term may come quickly, especially if migration inflows are predictable to investors. Possibly, skilled immigration boosts productivity and wages for many others. Around the averages, there are distributional effects. Among low-income "native" workers, the ones who stand to lose the most are those who most closely resemble new arrivals, in being immigrants themselves, being low-skill, being less assimilated, and, potentially, undocumented. But native workers and earlier immigrants tend to benefit from the arrival of workers different from them, who complement more than compete with them in production. Thus skilled immigration can offset the effects of low-skill immigration on natives and earlier immigrants.

---

[1] Thanks to George Borjas, William J. Carrington, and Giovanni Peri for data, code, and comments on earlier drafts; thanks to Holden Karnofsky and Ben Rachbach at GiveWell for comments too. Thanks to Colin Rust for a correction. Data and code for this review are at davidroodman.com/david/Stata files for Roodman review.zip.

## Introduction and summary

The immigration policies of wealthy nations could be prime leverage points for those seeking to influence US government policy for the global public good. "Observably identical" people—ones with the same education, age, sex, and so on—earn 2.0–15.5 times more in the United States than in poor countries such as Peru, Haiti, Egypt, and Yemen (Clemens, Montenegro, and Pritchett 2008). Allowing more low-income foreigners to work in wealthy nations could therefore raise their incomes more than almost any other step that governments or philanthropists can take.[1]

But there is a concern: Does allowing more immigration take jobs from people already in the labor force of the receiving country? Or does it depress their pay? This document performs due diligence on this question, especially as it pertains to low-skill workers in the receiving country, who can least afford to lose earnings. The document reviews the relevant economics literature to glean what is known about the impacts of immigration on employment and wages, as well as innovation and productivity, which affect earnings in the long run.

The debate over the economic impacts of immigration can be seen as a battle between metaphors. Is an economy like a pie, with a set number of jobs, so that one person's employment gain is another's loss? Or is it like a church congregation, which one person can join to experience communion and fellowship without costing anyone else the same?

To a first approximation the answer is clear: the latter. In wealthy nations at least, jobs and labor earnings are not in rigidly fixed supply. Jobs are relationships of exchange, forming and disintegrating every day. Like other relationships, their absolute number can expand with population. The arrival of new workers creates new possibilities for production even as it increases consumer demand for goods and services. Wealthy nations are wealthy precisely because of ability to squeeze value out of available resources, including human talent.

Still, that is only a first approximation. Real economies are not perfectly flexible. Nor are they guaranteed to meet the needs of the poor. So it is worth examining the evidence on the impacts of immigration on those already arrived, especially the poorest among them. This can have implications for immigration reform, perhaps including collateral policies to compensate for side effects.

As ever, the evidence base is not as sturdy as we would wish. Ironically, immigration policy is often arbitrary and even randomized (as in visa lotteries), which *has* allowed some high-quality measurement of impacts on *immigrants* (McKenzie, Gibson, and Stillman 2010; Clemens 2013b). Unfortunately, this randomness has not been as exploitable when assessing impacts on the receiving economy, because admissions have not been randomized across occupations, say, or cities.[2] Studies in one family attempt the next-best thing, exploiting *natural* experiments, and some are persuasive. The rest of the research is less experiment-like, for example, looking at correlations between wages and immigration flows across US cities over 20 years, and so must be taken with more grains of salt. Most of non-experimental studies reviewed here make a bid in the direction of natural experiments by *instrumenting*. But even when they aggressively check the instrumented results for robustness, it is hard to be sure that the strategy is

---

[1] Certain health interventions might rival the impacts of immigration, depending on the value placed on health and life. This is why the text compares immigration to other interventions in terms of income.

[2] Peri, Shih, and Sparber (2014b) ventures in this direction, but I argue below that it falls short.

working.[3]

Still, the available evidence paints a fairly consistent and plausible picture:

- There is almost no evidence of anything close to one-to-one crowding out by new immigrant arrivals to the job market in industrial countries. Most studies find that 10 percentage point increase in the immigrant "stock" as a share of the labor force changes natives' earnings by between –2% and +2% (Longhi, Nijkamp, and Poot 2005, Fig 1; Peri 2014, p. 1). Although serious questions can be raised about the reliability of most studies, the scarcity of evidence for great pessimism stands as a fact. The economies of destination countries largely appear flexible enough to absorb new arrivals, especially given time.
- The group that appears most vulnerable to competitive pressure from new low-skill migrants is *recent* low-skill migrants. This possibility is easy to miss when talking of the impacts of "immigrants" on "natives." Yet it stands to reason: a newly arrived Mexican with less than a high school education competes most directly with an earlier-arrived Mexican with less than a high school education.
- One factor damping the economic side effects of immigration is that immigrants are consumers as well as producers. They increase domestic demand for goods and services, perhaps even more quickly than they increase domestic production (Hercowitz and Yashiv 2002), since they must consume as soon as they arrive. They expand the economic pie even as they compete for a slice. This is not to suggest that the market mechanism is perfect—adjustment to new arrivals is not instantaneous and may be incomplete—but the mechanism does operate.
- A second damper is that in industrial economies, the capital supply tends to expand along with the workforce. More workers leads to more offices and more factories. Were receiving economies not flexible in this way, they would not be rich. This mechanism too may not be complete or immediate, but it is substantial in the long run: since the industrial revolution, population has doubled many times in the US and other now-wealthy nations, and the capital stock has kept pace, so that today there is more capital per worker than 200 years ago.
- A third damper is that while workers who are similar compete, ones who are different complement. An expansion in the diligent manual labor available to the home renovation business can spur that industry to grow, which will increase its demand for other kinds of workers, from skilled general contractors who can manage complex projects for English-speaking clients to scientists who develop new materials for home building. Symmetrically, an influx of high-skill workers can increase demand for low-skill ones. More computer programmers means more tech businesses, which means more need for janitors and security guards. Again, the effect is certain, though its speed and size are not.
- An important corollary of this last observation is that a migrant inflow that mirrors the receiving population in skills mix is likely to have the most benign effects. Especially once capital ramps up to match the labor expansion, a balanced inflow probably approximates a dilation of the receiving economy, with similar percentage increases in all classes of workers, concomitant growth in aggregate demand, and minimal perturbation in prices for goods, services, and labor. *In particular, one way to cushion the impact of low-skill migration on low-skill workers already present is to increase skilled immigration in tandem.*

Figures that capture a lot of these ideas come from Ottaviano and Peri (2012). Between 1990 and 2006 immigration contributed 13.2 percentage points of growth to the US stock of low-education workers (ones

---

[3] See the section, "On the econometric challenges of studying causality," in my review of the impacts of life-saving interventions on fertility (davidroodman.com/?p=422).

with no college) and 10.0 points to high-education workers (Ottaviano and Peri 2012, Table 1). As a result of this fairly skill-balanced inflow, they estimate, the average native-born worker's earnings rose 0.6%. However, the impacts of the inflow on earlier immigrants is estimated as much larger because it increased the number of immigrant workers in the US not by some 10%, but 125%. The added competitive pressure is calculated to have cut earlier immigrants' pay by 6.7%. For earlier immigrants with less than a high school education, the figure is 8.1%; and for those with a high school education but no college, it is 12.6%. (Ottaviano and Peri 2012, Table 6, col 7.)

If taken to approximate the truth, the impact estimates for earlier low-skill migrants pose a moral question for those pondering advocacy for more immigration. How should the economic harm to this relatively vulnerable group be weighed? That question is beyond the scope of this analysis, so I will touch on it only briefly. Some ways to answer it may be obvious: pure utilitarianism; conditioning support for more immigration on strengthening the social safety net even for undocumented residents already present. A less obvious approach is based on an argument for equal treatment of similarly situated people: that it would be unjust to protect recent immigrants, who have recently, say, tripled their incomes by moving from Mexico to the US (Clemens, Montenegro, and Pritchett 2008, Fig 3), by preventing other Mexicans from enjoying the same opportunity, especially if it would only cost the recent immigrants 20 points of their 200% gain.[4]

Before, I distinguished between randomized studies, natural experiments, and non-experimental ones. Another way to classify immigration impact studies is by the unit of analysis. Some are spatial: they looks at whether cities or states or provinces receiving more migrants over some period (whether or not because of a "natural experiment") experienced different trends in wages or employment for native workers, or patenting rates or productivity. Another influential set compares across groups defined by occupation or years of work experience and years of education. For example, do the education groups (or "skill cells") gaining more immigrants see wages rise or fall more for natives? Some studies split the data both ways at once—by place and by skill or occupation.

With those distinctions in mind, I have organized this review as follows. I look first at studies of impacts on wages and employment, then ones on innovation and productivity, factors that can drive wage growth in the long run. Studies in the wages-and-employment category are further organized by whether they make (convincing) cases to be exploiting natural experiments. Those not relying on natural experiments come first; and among them, the spatial one are reviewed first, then the skill cell–based ones.

# 1   Conceptual preliminaries

Our general question is what happens when new participants of a particular kind enter an industrial economy. These new participants will affect the receiving economy in two main ways: as consumers of output and providers of input. The symmetry in that simple observation suffices to suggest that the overall impact of immigration will be benign. The immigrants will compete with other workers, but they will also buy things, supporting increased production.

But immigrants' economic impacts as consumers and producers could differ distributionally, with net harm for certain vulnerable groups. Immigrants' economic impact as consumers is probably diffuse and similar in many ways to the distributional impact of the average consumer. Their shopping baskets contain high-and low-tech items, some made locally and some imported, some as tangible as bricks, some as intangible as credit. As participants in the labor market, however, immigrants' impacts could be more concentrated, hurting some workers and helping others. This is one reason most of the research focuses on labor market impacts.

---

[4] The 6.7% loss cited in the previous paragraph becomes a 20% loss expressed relative to pre-immigration earnings, since those are assumed one-third as large.

Accepting that focus, two dynamics seem key to shaping the impacts. The first has to do with how immigrant workers, as inputs, could complement or replace other workers in production. If new immigrants are interchangeable with most other workers, then the competitive effects of their arrival should be diffuse. The second dynamic has to do with how quickly the capital stock adjusts to the expansion of labor. The quicker the adjustment, the quicker the economy will achieve the growth made possible by additional workers, and the more transient the initial side effects. This section elaborates on both ideas.

## 1.1   Elasticity of substitution

An elasticity of substitution characterizes the relationship between two inputs to production—regular and premium gasoline, fuel and engines, natives and immigrants, labor and capital. It is a one-dimensional, spectrum concept. Understanding it requires this insight: distinct inputs can interact economically in two main ways. To the extent that they play the same role in production, like different grades of gasoline, they compete in the same product market, and greater supply of one can *reduce* demand for the other. But to the extent that they play *different* roles in production, greater supply of one can *increase* demand for the other, just as more engines in operation can raise gasoline demand.

At one extreme, two inputs can have an infinite elasticity of substitution, making them *perfect substitutes.* This does not quite mean that they are exact replacements for each other, but that they work like scaled versions of each other. A gallon of premium gas, for instance, might be equivalent in all ways to 1.1 gallons of regular unleaded. A driver could smoothly switch between the two without changing anything else, and obtain 10% more output from her engine per gallon of premium. The two types of fuel play the same role in the making of motion. Further, if premium cost exactly 10% more—as it will in the standard models of microeconomics—then the driver wouldn't care which she buys: she would get the same value per dollar.

At the opposite extreme is zero elasticity of substitution, or *perfect complementarity.* Like an engine and fuel, if two inputs are perfect complements, each is useless without the other. Flipping that around, expanded availability of each makes the other more valuable. And neither can substitute at all for the other. You can't fill your tank with engine.

Small-scale examples like these tend to produce extremes on the substitute-complement spectrum. But as one considers broader economic aggregates such as natives and immigrants, shades of grey appear. To an extent, for example, natives and immigrants compete for the same jobs, so that natives are hurt by competition from immigrants. But at the same time, the two groups complement one another: American restaurants commonly hire natives for the retail interface (waiting tables) and immigrants with little English for the kitchen. To this extent, expanded supply of one is good for the other. The arrival of people ready to work in the kitchens lets the restaurant business expand, raising demand for native wait staff.

Adding to the blurriness, degrees of substitutability vary over time. In the short run, labor and capital are mostly complements, as a strike can shut down a factory. In the long run, capital can substitute for labor through automation.

In principle, every input to a process like bicycle manufacture, viewed as a giant production process running from raw ores to finished goods, has an elasticity of substitution with every other input. With limited data, economists can only study a few elasticities at a time, so they focus on the ones most relevant to their research, estimate them from the data, and use the results to model how changes in the supply of economic inputs such as native and immigrant labor affect each other's prices.

The worst-case immigration scenario for native low-skill workers involves a certain combination of elasticities of substitution. First, native and immigrant low-skill workers would be perfect *substitutes*. Again, this would not mean that an immigrant is precisely as valuable to an employer as a native, but that the low-skill immigrants and natives can fill exactly the same roles in production, and so compete for the same jobs. Perhaps two immigrants could replace one native in a restaurant, at half the pay rate. Second, low-skill workers as a group, native and foreign-born, would in turn be perfect *complements* to skilled workers, such as restaurant managers. Under these circumstances, an influx of purely low-skill immigrants would increase competition for precisely the jobs low-skill natives can do. Yet for lack of additional managerial cadre—for lack of a parallel influx of high-skill workers—the restaurant industry would not expand. The pool of low-skill restaurant jobs would remain fixed, even as competition for them intensified. Native low-skill workers would be maximally squeezed.

The best-case scenario for low-skill natives is opposite: complementary between low-skill natives and low-skill immigrants (they mostly compete for different jobs) and high substitutability between low-skill workers as a group and other workers, so that the expanded pool of low-skill workers could push a bit into somewhat more-skilled jobs, diffusing the competitive impacts.

To investigate which scenario is most realistic, economists link the theoretical concept of elasticity of substitution with the concrete process of looking at correlations between wages and labor supply. The link is made by assuming that markets work reasonably well, in the following sense. In a perfect market, if a gallon of paint is perfect substitute for a quart, then it will *always* cost exactly four times as much as a quart. Increased production of quarts of paint might lower paint prices overall, but it will not break the iron rule of four. Thus the ratio in supply of quarts to gallons will have zero correlation with the ratio in price. At the opposite extreme, an increase in the supply of car engines in use *relative* to the supply of fuel could have a big, negative impact on the ratio of their prices: more engines per available fuel would mean higher prices for fuel relative to engines.

Perhaps a more intuitive way to think about elasticities of substitution is this: they tell us how integrated or segmented the labor market is.[5] The less interchangeable different kinds of workers are—the more they are like engines and fuel—the less they compete, and the more segmented the overall job market. An immigration-induced jump in the supply of one kind of worker—say, men under 30 without a high school degree—will create competitive pressure primarily within that group. The depressive effect will hit fewer workers' wages, but more acutely.

## 1.2   The supply of capital

At the macroeconomic level, labor and capital are partially substitutable and partially complementary. They are partially complementary because the most efficient way to make things almost always requires both labor and capital. To this extent, the arrival or workers should increase demand for capital, which will manifest initially as higher returns to investment. How quickly investors respond to that signal matters a lot for the overall economic impacts of immigration.

---

[5] But, as explained above, substitutability derives not only from the factor emphasized in this paragraph, heterogeneity among workers, but from complementary in production as well.

Again the possibilities lie on a spectrum between extremes. Suppose that as the labor pool expands, capital does not—at least, not for several years. There are more workers, but no more buildings, factories, and computers for them to work in and with. In this case, we would expect the glut of workers to depress pay for everyone. In the US economy, the accepted rule of thumb among modelers is that a sudden, unexpected 10% rise in labor supply reduces pay by 3% in the short term, which by definition is the period in which capital fails to respond to a labor expansion.[6] The elasticities of substitution discussed in the previous subsection would still govern the relative distribution of impacts, that is, which workers' pay would fall more than 3% and which less.

But in the long run, history assures us, the capital supply *will* to a substantial degree expand to exploit the investment opportunities created by the arrival of workers. After all, the US labor pool has doubled many times since 1776, and capital has more than kept up. If capital supply is only a transient constraint on the economic absorption of workers, then we should expect the arrival of more workers to have *zero* average impact on wages in the long run. There will be more workers, more buildings, more factories, more computers, all producing things and selling labor at about the same prices—to more consumers.

Notice that the first scenario is short-term and the second long-term. So the main practical question is not whether one scenario is right and the other wrong, but how short the short term is. How quickly do investors take advantage of an increment to labor? An ethical question also arises: Given our best estimate of adjustment speed, how much should we worry about short-term harm if the long-term harm is minimal?

Unfortunately, the empirical question is hard to answer, for the same reason that the causes of economic growth are hard to determine. The nation becomes the unit of observation, and so many things vary simultaneously within and across nations that cause and effect are very hard to study statistically.

Meanwhile, it is worth keeping in mind that for reasons outside of the conventional labor-capital dynamic, immigration could raise average wages. Some studies suggest that immigration, particularly of skilled workers, accelerates technological innovation, which is the basis of economic growth, and can lead to higher pay for large swaths of the workforce. Immigrants are notably important in computer-related industries.

My sense is that in the flexible US economy, the long term capital response comes pretty quickly. Most wealthy nations have imported a lot of capital as well as labor in recent decades, indicating significant capacity for smooth absorption of immigrants into the economy. Immigration increases of the sort that could occur under realistic policy change would add up over time but be small and predictable on an annual basis. Even a doubling of US immigration from 1 million to 2 million/year would expand the US labor force by only 0.65%/year.[7] And if the increased arrivals of new workers and consumers were predictable, investment patterns might effectively *anticipate* each year's expansion, projecting this year's growth based on last year's. (The macroeconomic tradition of "rational expectations" asserts as much.) The comparatively benign "long run" would arrive immediately.

---

[6] The US economy is usually taken to be Cobb-Douglas, with total output $Y = K^\alpha L^{1-\alpha}$ where $K$ is capital and $L$ is labor. A perfectly flexible, profit-maximizing economy will settle into an equilibrium where the pay to each factor equals its marginal value. In particular, with $K$ fixed, wages $w$ will be $\partial Y / \partial L = (1-\alpha) K^\alpha L^{-\alpha}$. The share of total earnings in economic output will then be $Lw/Y = L(1-\alpha) K^\alpha L^{-\alpha} / K^\alpha L^{1-\alpha} = 1 - \alpha$. Economists observe that over many decades, US workers have taken home about 70% of national earnings (Ottaviano and Peri 2012, Fig 1), suggesting that $\alpha \approx 0.3$. So, under the model, the elasticity of wages with respect to labor supply will be $\partial \ln w / \partial \ln L = \partial \ln\big((1-\alpha) K^\alpha L^{-\alpha}\big) / \partial \ln L = \partial[\ln(1-\alpha) + \alpha \ln K - \alpha \ln L] / \partial \ln L = -\alpha \approx -0.3$.

[7] The US labor force is 155 million workers. cia.gov/library/publications/the-world-factbook/geos/us.html.

On the other hand, while the hypothesis for a productivity bump from immigration is plausible, the evidence is weak (see below). As a result, the best anchoring assumption for the long-term response of national average wages to total labor supply looks to be not positive but zero (Card 2012, p. 211–12).

## 2  Spatial studies that do not exploit natural experiments

Cross-city and cross-region studies are akin to cross-country studies, which, as I explained in my [review of the impacts of mortality on fertility](), are now in disrepute. But analyzing *within* rather than across countries is more reliable because it shortens the list of potential confounders, factors left out of a model that could be simultaneously influencing the variables of interest. Iowa City is more like San Diego than Zimbabwe is like China because Iowa City and San Diego have long been subject to similar policies and economic conditions. To the extent that these deep traits are the same across cities, they cannot explain differences in immigration or wages or unemployment.

Most spatial studies have found only small impacts of immigration on receiving geographies. Summarizing the literature to that time, Card (2001), p. 23, writes, "Typically, a 10-percentage-point increase in the fraction of immigrants (roughly the difference between Detroit and Houston) is estimated to reduce native wages by no more than 1 percentage point." (Studies finding low impacts include Grossman (1982), Altonji and Card (1991); LaLonde and Topel (1991). On the other hand, Schoeni (1996) finds effects closer to 1:1.)

However, spatial studies have been criticized for several shortcomings that could cause them to underestimate impacts (Chiswick 1991, p. 627–28; Borjas 1994, p. 1699). A first concern is that to lump an Indian computer programmer together with a Mexican farm worker is to obscure the texture of immigration's impacts. Immigrants must be distinguished from each other at least by education level. The remaining concerns are, econometrically, of the kind mentioned just above—about potentially important confounders that are ignored. First, causality may go the other way, and with opposite sign. Maybe the dominant story is not that immigrants arriving in New York reduce wages there, but that a strong New York economy is lifting wages and attracting more immigrants. This positive relationship would not preclude a negative impact of immigration on wages, but it would obscure it. Second, if immigrants depress wages in a city, then some workers may move to other cities, depressing wages there too. And if wages fall everywhere at once, then comparing across cities will expose no effect. In other words, the labor market is open and national, not closed and local. Third, the same masking could occur through internal trade. A migrant influx could depress wages in a city's labor-intensive, goods-producing businesses. Those businesses would lower their prices to undercut competitors in other cities, forcing the competitors to cut pay too. Again, the wage drop would be national, so cross-city comparisons would wrongly register no effect. Finally, if industrial patterns shift, then short-term depressive impacts could be missed in studies that look long term. If the arrival of migrants in certain cities depresses wages, it may attract businesses such as clothing manufacture that are sensitive to low-skill labor costs, or cause many industries to reverse automation (Lewis 2011). Added demand could eventually bring wages back to the old level.

Perhaps the most sophisticated spatial study is Card (2001), which works to parry all these charges.

## 2.1 Card (2001), "Immigrant Inflows, Native Outflows, and the Local Market Impacts of Higher Immigration," *Journal of Labor Economics*

In truth, though I class this study as spatial, its unit of observation is not the city, but the city–occupation group combination. High-skill workers in Phoenix, say, are compared to low-skill ones in Boston and medium-skilled ones in Miami.

In particular, Card divides occupations into six broad groups roughly stratified by skill level and pay. Group I, for example, consists of "laborers, farm workers, and low-skill service workers," while Group VI is "professionals and technical workers." Card views immigration into a city as a set of differentiated increases in the supplies of people statistically predicted to find employment in each group. Nationally in the late 1980s, for example, 39.2% of immigrants from Cuba were statistically expected, based on such traits as age, sex, and education, to work in Group I and only 8.3% in Group VI. For India, Pakistan, and Central Asia, the numbers are 21.9% and 28.0%.

Notably, the trait of being a new or earlier immigrant *only* enters the model in this way. Despite differences in linguistic and cultural assimilation, not to mention legal status, an immigrant classed as Group I is viewed as economically interchangeable with a native in Group I. Immigration flows manifest as increases in a city's stock of workers in each of the six occupation groups.

Card's empirical questions take the form: in 1990, in cities and occupation groups that since 1985 gained more people whose demographic traits statistically suit them for those groups, are employment and wages lower? How does the answer differ between natives and earlier immigrants, and between men and women? These differentiations answer the first criticism of spatial studies cited above, about treating immigrants as homogeneous.

Card also takes on the endogeneity-related criticisms. In response to the first, that natives will undermine cross-city comparisons by fleeing, and thus spreading, low wages, Card checks whether that happens. According to his analysis of the 1990 US census, it does not (Card 2001, Table 4). If anything, cities that attracted more international immigration also attracted net domestic immigration. And to address the challenge that the same dissipation of apparent impacts could occur through intercity trade, Card studies short-term effects—from immigration during 1985–90 to employment outcomes on census day in 1990—arguing that the trade effects could take a decade to kick in.[8,9]

Finally, Card takes common steps to reduce endogeneity: adding controls and instrumenting. Looking across two dimensions of variation at once, place and occupation, allows Card to control for fixed effects associated with each dimension.[10] For example, if there is something about New York that simultaneously attracts immigrants and raises wages in all occupation groups proportionally, this confounding is expunged.

As for instrumenting, Card observes that new immigrants tend to go to the same cities as old ones, and where old ones went cannot be reversed-caused by labor market outcomes today. Card's most rigorous regressions look at whether labor supply expansions attributable to historical immigration patterns are correlated with higher unemployment or lower wages for those already here.

---

[8] Thus the lag between immigration and measured impact ranges here between 0 and 5.25 years, averaging under 3.
[9] Studying Massachusetts after an unemployment spike circa 1990, Blanchard and Katz (1992, Fig. 7) find labor market adjustment to take 5–10 years.
[10] On "fixed effects" see the introduction of the section, "Cross-country panel studies," in my review of the impacts of life-saving interventions on fertility (davidroodman.com/?p=422).

But I find this econometric step the weakest: dominant destinations for migration, such as New York, Florida, Texas, and California have differed systematically from the rest of the country for decades. They may possess deep, long-term traits that support economic growth and attract immigrants, so that a potentially misleading positive correlation between job market conditions and immigration still obscures the harm of immigration for some.

That said, introducing Card's instruments produces more negative results, suggesting that the instruments are doing their job, if perhaps imperfectly, removing the positively signed bias of reverse causation. Without instruments, a 10% increase in the number of people in a given city and occupation group is associated with a 0.28 percentage point drop in employment among men in that group (0.45 points among women). The drops just for employment of earlier immigrants are milder, at 0.19 and 0.23 points. (Card 2001, Table 6, row A.) After instrumenting, the estimates are much larger: 2.02 and 0.81 points for male and female natives, 0.96 and 1.46 points for earlier immigrants (Card 2001, Table 6, row D). But the native–earlier immigrant differences are not statistically significant.

It bears emphasizing that the results for earlier immigrants do not arise from directly observing their employment experience. What is observed is that in given cities and occupation groups, employment is higher or lower by certain amounts, which are assumed the same for natives and earlier immigrants in the groups. Yet since natives and earlier immigrants are distributed differently across these groups, they are estimated to experience different employment changes on average. Some studies reviewed below assess impacts on earlier immigrants more directly, and thus more reliably.

The pattern of estimated impacts on *wages* is less consistent than that for employment: for a 10% labor supply increase, the wage declines are 0.99% for male natives, 2.5% for female earlier immigrants, indistinguishable from zero for other groups (Card 2001, Table 7, row 4).

Overall, while there is reason to worry that the study still underestimate the impacts (the instruments are not perfect), the results suggest that a 10% labor supply increase from immigration reduces employment and wages among those already in the labor force on the order of 2% in the short run. This finding of modest impact corroborates the large majority of spatial studies.

## 2.2   Cortes (2008), "The Effect of Low-Skilled Immigration on U.S. Prices: Evidence from CPI Data," *Journal of Political Economy*

Cortes (2008) is billed as a cross-city and cross-state examination of the impact of immigration on *prices*. Prices might seem distant from *wages*, but not those Cortes focuses on, which are for services such as such as childcare, housekeeping, gardening, dry cleaning, and shoe repair, the costs of which are dominated by low-skill wages. And the paper directly examines impacts on wages too.

Like Card 1991 and many other spatial studies, Cortes (2008) instruments the immigration inflow with a prediction of that inflow based on past immigration patterns. And as in Card (2001), while the instrument is not beyond challenge, introducing it shifts the results in a direction that makes sense. Without the instrument, the correlation across US cities between low-skill immigration and prices for low-skill-intensive services is mildly negative. With the instrument, the apparent impact of immigration on the studied prices rises three- to five-fold (Cortes 2008, Table 5). This suggests, plausibly, that low-skill immigrants move to cities where the prices for (and thus wages in) low-skill-intensive services are higher, which positive correlation statistically masks any downward pressure the immigrants put on wages upon arrival. Evidently, instrumenting removes some or all of this effect.

Also like Card 1991, Cortes checks whether the number of natives working in a city falls as immigrants arrive, and finds that it essentially does not. According to her regressions, each additional migrant to a city in 1980–2000 expanded the local labor force by 0.8–1.3 workers, net (Cortes 2008, Table 4, row 1).

The pattern of results is persuasive in another way: the association of low-skill immigration with prices is weaker for services that require more skill or more natives, and disappears altogether for traded goods whose prices are largely set by world markets (Cortes 2008 Table 6). This pattern is consistent with immigration lowering just certain prices and is hard to explain with other causal stories.

For a bottom-line figure, Cortes (2008), p. 401, estimates that in a representative US city, each 10% increase in the number of low-skill immigrants cut prices of low-skill-immigrant-intensive services by almost 2% (Cortes 2008, p. 401.). She corroborates with regressions estimating similar drops in wages for low-skill earlier immigrants, as well as for natives who resemble them in being Hispanic and non-English speaking. Wages for most low-skill natives, on the other hand, show no change. (Cortes 2008, Table 8 & 10.) Since the association with the wages of earlier immigrants is directly measured, not inferred as in Card (2001), the finding of greater impact on earlier immigrants is more reliable than Card (2001)'s more ambiguous findings on that score. The Cortes reading supports the belief that low-skill natives and immigrants are to a degree complementary in production, so that the competitive pressure of new immigrants concentrates on earlier ones.

## 2.3   Peri, Shih, and Sparber (2014a), "Foreign STEM Workers and Native Wages and Employment in U.S. Cities," working paper

This study also fits firmly in the spatial tradition, comparing across US cities and instrumenting current migration patterns with past. It distinguishes itself by focusing on science, technology, engineering, and mathematics (STEM) professionals.

The study is clear-eyed about the difficulty of determining causality in this set-up and takes aggressive steps to test the robustness of its approach. To take one example of a concern combatted, perhaps cities with large stocks of college-educated workers have been more economically successful for decades, driving up wages and attracting more-skilled immigrants for a long time. This causal story would invalidate past immigration as an instrument. To check this possibility, Peri, Shih, and Sparber (2014a), p. 17, add a control for the city's growth in college-educated natives; it does not undermine their results.

Cutting against the study is the size of the estimated impacts. A 1 percentage point increase in a city's growth of foreign STEM workers leads to a stunning 7–8% wage increase for natives (Peri, Shih, and Sparber 2014a, p. 3). As a result, immigration of STEM workers can explain 30–50% of all US productivity growth between 1990 and 2010! (Peri, Shih, and Sparber 2014a, p. 4.)

Such results make it hard for me to dismiss the possibility that reverse- or third-variable causation is strongly at work. It is much easier to believe, for example, that a 7–8% wage increase for natives led to a 1 percentage point increase in growth of STEM worker immigration.

## 2.4 Peri, Shih, and Sparber (2014b), "The Effects of Foreign Skilled Workers on Natives: Evidence from the H-1B Visa Lottery," working paper Partnership for a New American Economy (PNAE) 2014, *Closing Economic Windows: How H-1B Visa Denials Cost U.S.-Born Tech Workers Jobs and Wages during the Great Recession*

Peri, Shih, and Sparber's second study attacks the same question as the first, but by trying to exploit a natural experiment, which is the random allocation of H-1B visas through a yearly government lottery. But I place the paper in this section because it appears not to be a natural experiment study in practice.

Under the H-1B program, started in 1990, US employers wanting to hire skilled foreigners must apply for a visa for each slot they want to fill. The visas last three years and are renewable once. In 2004, the national visa cap fell back from the temporarily elevated level of 195,000/year to 65,000/year. As a consequence, in 2007 and 2008, an average 88,693 visa applications/year, or 57.5%, were denied. Since the lottery was national, the denial percentage varied randomly by city. Peri, Shih, and Sparber study how this arbitrary allocation differentially affected pay for native computer workers in 2009–11.

The prospect of exploiting "genuinely random variation" tantalizes. Unfortunately, in this case it also overpromises. The problem is that the treatment variable is the product of two factors, only which of is random—and it may be secondary mathematically. We can write the treatment variable like this:

$$\text{City's visa denial rate} = \frac{\text{\# of computer profession visas requested}}{\text{\# of computer professionals already in city}} \times \text{\% of visa requests denied}$$

Only the second term is random, and its value may not vary much. Why not? All applications faced the same denial probability and all cities had about the same share denied.

In fact, by the law of large numbers, the more visas a city requested the more its denial rate converged to the national average of 57.5%. The city with the largest shock, for example, was Trenton, NJ, which had 2,280 computer programmer visas denied on average in 2007–08, against an existing computer-related workforce of 6,148 (PNAE 2014, Table 1). The national denial rate of 57.5% implies that Trenton firms requested about 4,000 visas. Like a presidential poll of 4,000 people, the "margin of error" in Trenton's lottery experience—the deviation of its denial rate from the national average—is pretty small: ±1.5%. That is, there is a 95% chance that Trenton's denial rate was between 56% and 59%. (The actual rate cannot be determined from the figures displayed in the study reports.)

In general, a city's margin of error is approximately $1/\sqrt{\text{\# of visa applications}}$.[11] For the average city, the margin of error is 3.8%, which is 6.6% of the average denial rate of 57.5%.[12] So most cities' denial rates clustered in a narrow band around the average.

Meanwhile, the first term in the above formula varies much more in proportion to its mean. It runs as high as 4,000/6,148 = 65% for Trenton to, presumably, close to zero for many metropolitan areas.[13] It averages 7.0%.[14] As result, this nonrandom variation appears to dominate the seemingly randomized experiment.

---

[11] If $x_i$ is the 0/1 *denial* outcome for visa application $i$, its variance is $E[x_i^2] - E[x_i]^2 = .575 - .575^2$. The variance of the average of $N$ uncorrelated draws $x_i$ is $(.575 - .575^2)/N$, for a standard error of $\sqrt{(.575 - .575^2)/N}$. 95% confidence intervals are based on ±1.96 of these standard errors, $\pm 1.96 \times \sqrt{(.575 - .575^2)/N} = \pm 0.97/\sqrt{N}$.

[12] 88,693 denials + 65,000 acceptances per 236 metropolitan areas = 651 applications/area. $0.97/\sqrt{651} = 0.038$.

Peri, Shih, and Sparber find that by stunting the growth of technology companies, the 177,386 visa denials in 2007–08 cost US cities some 231,000 tech jobs for *natives* by 2010. And it reduced wages for natives with computer-related jobs by 4.9% (PNAE 2014, p. 5).

Since the econometrics do not buttress their causal story the way true randomization would, we must think critically about the results. Could alternative theories explain the association between visa denials by city and lower employment and wages for natives? I think so. The nonrandom component in the visa denial equation above depends on the difference between the number of people a city's computer-related companies want to hire and the number of potential workers who are available locally. Since local technology industry growth can be path dependent—it happens more where it has already happened, and computer workers may move to areas with a reputation for technology leadership—the local gap between computer workers needed and computer workers available could be significantly higher in places where the tech industry is attempting to grow above its historical rate. These could also be places where demand for computer professionals has historically been low, reducing their employment and wages (and the local cost of living) today. This story could explain the association between greater demand for H-1B visas, and thus greater exposure to visa denials, and lower employment and wages among natives.

Meanwhile, apparently because of the reliance on the perceived randomized experiment, Peri, Shih, and Sparber (2014b) does not conduct the sort of aggressive robustness tests found in Peri, Shih, and Sparber (2014a), which might deprecate or rule out competing theories.

## 2.5   Summary: spatial studies that do not exploit natural experiments

Card (2001) provides credible corroboration of small, negative overall labor market impacts while significantly blunting the main criticisms of spatial studies. As for differences between natives and earlier immigrants, Cortes (2008) is most compelling because her estimate are based on more direct observations of these two groups. Overall, this sub-literature suggests that low-skill natives have little to lose from low-skill immigration, but that earlier immigrants pay a price on the order of 2% for each immigration-induced 10% increase in immigrant labor supply.

# 3   Skill cell studies that do not exploit natural experiments

Another major stream in the literature works purely with the variation across skill groups. Here, the unit of observation is the "skill cell," the set of people across an entire country who have a certain degree of education and/or a certain length of experience in the workplace—these traits being major determinants of skill level and earnings. The empirical question becomes whether, on a nationwide basis, wages and employment for natives and earlier immigrants move differently in skill cells receiving more new immigrants.

One attraction of the skill cell approach is that it surmounts some of the criticisms of spatial studies enumerated earlier. Recall that one criticism is that people can easily move from city to city, making intercity comparison potentially misleading. The equivalent in skill cell studies would be for people to suddenly gain more years of work experience, which they can't, or earn a GED or college degree, which they can do only with difficulty. To the extent that people are stuck in their cells, the wage and employment effects of new arrivals are less apt to be masked by compensating shifts on the part of those already in the domestic labor market. (On the other hand, Card (2001) and Cortes (2008) didn't find much *spatial* movement either.)

---

[13] Data are only reported for the 20 of the 236 metropolitan areas with the largest shocks (PNAE 2014, Table 1).
[14] 153,693 applications against a computer-related workforce of 2,188,258 (PNAE 2014, Table 1, last row).

However, skill cell studies have an important weakness. Because they view the national economy as integrated, not as a set of discrete metropolises or states, it is hard for them to estimate the overall relationship between labor supply and wages. They cannot ask whether wages were depressed in cities where the workforce grew faster—a question that might be answered by comparing data for 100–200 cities. They can only ask, when the national labor supply went up, did wages rise or fall? In trying to answer this question, there would be only one, national data point per time period to work with—and many theories competing to explain it. Moving from decennial census data to annual data would not help much, because successive years are not statistically independent. The US in 1960 was a lot like the US in 1959.

As a result, in interpreting skill cell studies, it is essential to distinguish between relative effects and absolute averages. One must bring *to* the studies an assumption about absolute average wage impacts—for which, as argued earlier, zero is probably a good guess especially after a few years to give investors time to adjust. The skill cell regressions can only speak to redistribution, to whether the impacts of immigration are concentrated or diffuse. They can speak, for example, to whether immigration increases inequality by widening the pay gap between those who have spent time in college and those who have not.

The skill cell literature has been particularly disputatious. Most of the arguments are over which elasticities of substitution should figure in economic models—those between natives and immigrants, perhaps, or the less- or more-educated, or the less- or more-experienced—and over how best to estimate them. Researchers have also differed in their interpretive emphasis of the short term versus the long term, the latter being more optimistic in its allowance for capital adjustment.

## 3.1 Borjas (2003), "The Labor Demand Curve Is Downward Sloping: Reexamining the Impact of Immigration on the Labor Market," *Quarterly Journal of Economics*
Borjas (2014), *Immigration Economics*, chapter 5

Viewing the economy as a giant production process, every input has an elasticity of substitution with respect to every other. In order to tame that complexity, when labor economists study the impacts of immigration, they work with a few abstract categories. There is capital, which embraces all investment in tangible and intangible assets, and there is labor. Labor is typically subdivided according to the education level of workers and their years of experience, on the idea that learning occurs both before after entering the work force. Borjas (2003), for example, distinguishes four education groups—people who did not complete high school, who completed just high school, who had some college, or who completed college—and eight experience levels—1–5 years, 6–10 years, …, up to 36–40. Borjas (2014) further divides "completed college" into "college graduate" and "post-graduate."

Even these $1 + 4 \times 8 = 33$ or $1 + 5 \times 8 = 41$ input categories produce unmanageable complexity, for among them are 528–820 possible elasticities of substitution.[15] To simplify matters, economists may impose additional assumptions by creating a hierarchy (Card and Lemieux 2001). Borjas (2003) introduced

---

[15] From 33 categories, $32 \times 33/2 = 528$ $32 \times 33/2 = 528$ pairwise selections are possible. For 41, 820 are possible.

this idea to the study of immigration.

As already hinted, Borjas starts by taking the most fundamental split as being between labor and capital. The substitutability between labor and capital is one influence on wages, and like the overall relationship between labor supply and wages, the interplay between labor and capital is hard to determine from national-level data. So an assumption about it too must come from outside of skill cell empirics.

Within labor, Borjas's next split is by education level. The degrees of substitutability among the four or five education groups are assumed to be all the same—so high school drop-outs compete with or complement high school graduates precisely as much as they do with college graduates.[16] In the same way, education groups are subdivided into experience subgroups, which are again assumed to compete equally among each other.

These simplifications no doubt do violence to reality. But they allow economists to estimate complementarity effects—how low-skill immigration can benefit high-skill workers and vice versa. In contrast, some studies ignore these positive complementarities across skill groups, looking only at the negative competitive effects within groups.

All this structure reduces Borjas's task to estimating just two elasticities of substitution: across education groups; and, within education groups, across experience groups. This he does using Census Bureau data for 1960–2000 (updated to 2010 in Borjas 2014). Concretely, for each combination of education level, experience level, and census round between 1960 and 2000 or 2010, Borjas averages wages and person-weeks worked nationwide. This distills a massive database down to just 160 data points (240 in Borjas (2014)).[17]

The analysis proceeds in several steps that are too complex to fully explain here, which have the effect of climbing the hierarchical structure stepwise from the bottom. The first step estimates the bottom-most elasticity of substitution, between different experience groups within education groups, e.g., between college graduates with 6–10 years of work experience and college graduates with 11–15. It regresses wages for each education-experience group on total hours worked by that group in the given census year (both variables in logarithms), while including dummies dictated by the model to capture certain fixed effects. (The appendix below has more detail.)

For all the conceptual superstructure, the standard worries about reverse and third-variable causation pertain, as emphasized in Borjas, Grogger, and Hanson (2012). The assumption is made that labor supply and wages are linked though only one causal channel, running from labor supply to wages.[18] So while we should not dismiss the results out of hand, we should recognize that robust theoretical motivation does not inoculate the regressions against the usual econometric worries.

Borjas (2003) estimates the elasticity of substitution between different experience classes within an education class at 3.5, and that between education classes at 1.35 (Borjas 2003, eqs. 16 & 17).[19] By the standards of this literature, these numbers are rather low. (They can go as low as 0 and as high as infinity.) They indicate that workers of different education and experience levels do not easily substitute for each other. Rather, to a significant extent, differently skilled workers seem to compete in different labor markets, so that the competitive wage effects of an influx of inexperienced, minimally educated immigrants will tend to concentrate among already-arrived workers of the same skill class. Workers in complementary, more-skilled groups will tend to benefit, just as would-be restaurant managers can benefit from the appearance of more kitchen workers.

---

[16] Ottaviano and Peri (2012, p. 184), experiment with inverting the hierarchical relationship of education and experience and allowing different substitutability between high drop-outs and graduates on the one hand and those with some college and full college on the other (their "Model B"). They find that the data oppose the first change and support the second.

[17] The latter gains an education category, as noted, and 2010 census data.

[18] However, because the included dummies for each survey-education combination and each education-experience combination can soak up any endogeneity that is fixed within each of either of those groups, some competing causal stories are countenanced and removed.

Borjas (2003) uses these results to estimate the impacts of actual US immigration during 1980–2000 (see Table 1). This influx was barbell-shaped—largest at the least- and most-skilled ends. Importantly, and somewhat oddly, Borjas estimates the short-term impacts of this long-term development. That is, he assumes that the capital stock does not expand at all in response to the new workers. Since the inflow equaled about 10.7% of the US workforce, overall wages fall by assumption by $0.3 \times 10.7\% = 3.2\%$. (The subsection above on "The supply of capital" discusses the origin of the 0.3 multiplier.) Around that average of –3.2%, the drop was greatest in the education classes receiving proportionally more immigrants. Workers at the ends of the education spectrum experienced more added competition and benefited less from complementarity with new arrivals in other education classes; their wages are estimated to have fallen more than –3.2% (See first two columns of Table 1.)

Borjas updates this analysis in a June 2014 book (Borjas 2014). He appears to make two major changes: adding 2010 census data, and simulating full capital adjustment as well as no capital adjustment. Perhaps because of the additional data, the elasticity estimates rise: 6.7 instead of 3.5 across experience classes, 5.0 instead of 1.35 across education classes. This implies that the wage impacts of new arrivals are more evenly spread across all workers. Meanwhile, allowing capital to fully adjust adds 3.2% to all wages. As a result, the 8.9% drop for the least-skilled in Borjas (2003) becomes a 6.2% or 3.1% drop in Borjas (2014) (see last two columns of Table 1).

*TABLE 1. ESTIMATES OF THE WAGE IMPACTS BY EDUCATION CLASS OF US IMMIGRATION*

---

[19] The elasticity of 1.35 suggests, for example, that if immigration swelled the ranks of workers without a high school degree by 1% relative to those with, then the wages of below-high school workers would fall by $1/1.35 \times 11/1.35 \times 1$ relative to above–high school ones.

| | Immigration-induced rise in hours worked | Estimated wage impact | | | | |
|---|---|---|---|---|---|---|
| | | Borjas 2003 | Ottaviano & Peri 2012 | | Borjas 2014 | |
| Years | 1990–2010 | 1980–2000 | 1990–2006 | | 1990–2010 | |
| Capital stock adjusts to more labor? | | No | Yes | | No | Yes |
| Nativity of workers affected | | Native | Native | Foreign | Native | |
| | | | | | | |
| Education level | | | | | | |
| No degree | +25.9% | –8.9% | +1.7% | –8.1% | –6.2% | –3.1% |
| High school degree | +8.4% | –2.6% | +0.6% | –12.6% | –2.7% | +0.4% |
| Some college | +6.1% | –0.3% | +1.2% | –2.2% | –2.3% | +0.9% |
| College graduate | +10.9% | –4.9% | 0.0% | –5.5% | –3.2% | –0.1% |
| Post-graduate | +15.0% | | | | –4.1% | –0.9% |
| Overall | +10.6% | –3.2% | +0.6% | –6.7% | –3.2% | 0.0% |

Table shows estimated impacts of immigration on the wages of workers already in the US labor force.
Borjas 2003 and Ottaviano & Peri 2012 group together college graduates and post-graduates.
Sources: Borjas 2003, Table IX; Ottaviano & Peri 2012, Table 6, col 7; Borjas 2014, Table 5.3.

## 3.2 Ottaviano and Peri (2012), "Rethinking the Effect of Immigration on Wages," *Journal of the European Economic Association*

Ottaviano and Peri (2012) (OP) elaborate on Borjas (2003) in several ways. They update the data from 2000 to 2006 (between censuses, the Census Bureau conducts other representative surveys). They allow the substitutability between the lower two education categories (those with or without a high school degree) to differ from that between the better-educated (some college or college graduate), the intuition being that today high school drop-outs can fill about the same economic roles as high-school graduates while the same cannot be said for community college graduates and PhD's. Most important, OP expand the Borjas (2003) hierarchy at the bottom, splitting between native and foreign-born workers.

The significance of the native-immigrant split can be seen in the restaurant example: by not splitting natives and foreign-born, Borjas (2003) treats them as perfect substitutes: immigrants can wait tables too, if perhaps not quite as efficiently as natives. But perhaps, OP suggest, that is wrong. Perhaps new immigrants sometimes play distinct roles in production, for lack of linguistic and cultural assimilation and possibly legality. In this case, the competitive effect of new immigrants will concentrate to some degree among earlier immigrants, and natives will be partly shielded. Natives might even benefit on net because of complementarity: cheaper kitchen workers means more restaurants, which means more jobs for wait staff.

Ottaviano and Peri (2012), p. 156, estimate the elasticity of substitution between immigrants and natives at about 20. This means that a 20% increase in immigrants per native reduces the pay of immigrants relative to natives by 1%. Rerunning the regression after splitting the sample at the high school graduation line produces a native-immigrant elasticity of 11.1 among those only a high school degree or less, and 33 for those with more; the latter estimate is statistically indistinguishable from infinity (Ottaviano and Peri 2012, p. 185). This suggests, plausibly, that educated immigrants and natives are more interchangeable than their less-educated counterparts. Probably educated immigrants arrive better able to function in American language and culture; and probably any limitations in this regard matter less to their employers if, for example, the immigrants code well.

Even the native-foreigner elasticity of 11.1 within the less-educated, though not Borjas (2003)'s assumed value of infinity, is pretty high: evidently low-skill immigrants visit only a bit more competition on earlier migrants than on natives. Perhaps the difference is small because OP's immigrant group contains all foreign-born workers, including those who have had decades to assimilate to the point where they differ little as workers from natives. At any rate, immigration has so dramatically expanded the ranks of foreign-born laborers in the US in recent decades that even this low incremental impact adds up. According to the OP data, hours worked by foreign born without a high school degree increased 125% between 1990 and 2006.[20] The result, in the OP assessment, was relatively intense cumulative pressure on pre-1990 arrivals—roughly, a 125%/11.1 drop in their earnings relative to new arrivals.

In OP's simulations, excerpted in Table 1, the resulting native–foreign-born split is clear. They estimate that natives experienced no wage falls, but that earlier immigrants saw drops of 2.2–12.6%. The impact estimates average zero across all groups because capital is assumed to adjust to labor expansion rather than being a bottleneck to labor absorption as in Borjas (2003).

Table 1 contains an irony. While OP are generally seen as the immigration optimists in their debate with Borjas and colleagues, their results are more negative in the sense of indicating more concentrated harm on a vulnerable group, previous immigrants. OP see *worse* impacts on inequality.

## 3.3   Borjas, Grogger, and Hanson (2012), "Comment: On Estimating Elasticities of Substitution," *Journal of the European Economic Association*

Borjas, Grogger, and Hanson (2012) (BGH) replicate and critique Ottaviano and Peri (2012). The final versions of both papers appear in the February 2012 issue of the *Journal of the European Economic Association*, capping a long back-and-forth between the parties. Evidently Borjas and colleagues perceived OP, with their optimism about the impacts on *native* workers, as a misguided dissent from Borjas (2003).

I have run and pondered the code behind OP, BGH, and the Borjas (2014) update to Borjas (2003). I think that the BGH critique raises some good questions about the fragility of OP's novel results. But the arguments are also overdrawn. And to the extent they hold, they damage not just OP, but the whole skill cell approach initiated by Borjas (2003). The appendix provides a full, technical discussion.

At base, the two sides differ on two important questions, and on both I find the OP view more realistic. The first is the issue already discussed about how fast the capital stock responds to labor expansion. Borjas (2003) accentuates the pessimistic short-term perspective, and BGH endorses it through silence on the issue. OP focus on the long-term, in which capital adjusts, and which seems more relevant when immigration is steady. (Borjas (2014) is more even-handed, doing both.)

The second issue is the reliability of the OP finding of modest native-immigrant complementarity, which implies better outcomes for natives and worse for earlier immigrants.

The strongest BGH criticism launches from a point I explained in reviewing Borjas (2003), that for all the theoretical superstructure, one can debate whether the econometrics attack endogeneity aggressively enough with controls and instruments. BGH observe that adding dummies for a large set of fixed effects to the OP regressions greatly widens the confidence intervals around the estimate of native-immigrant substitutability, to the point where one can no longer reject perfect substitutability. After all, maybe new

---

[20] Figure obtained by loading the OP Stata data file "supply_groups_60_06_specification_1.dta" (webcitation.org/6RMHAjv1M) and typing "table year if year==1990 | year==2006, c(sum howo_ma_for sum howo_fe_for)", which produces 1990 and 2006 total hours worked by foreign-born males and females.

immigrants put as much competitive pressure on natives as on earlier immigrants.

I take that point seriously, but also see issues. First, in imposing that standard of rigor on OP, BGH do not acknowledge that the OP regressions are *already* more rigorous than the corresponding ones in Borjas (2003) (and now in Borjas (2014), ch. 5). Second, for reasons laid out in the appendix below, in augmenting the OP regressions in order to rule out other causal theories, BGH abandon the original hierarchical construct. Imposing the BGH standard of rigor on the Borjas works and OP robs all the regressions of theoretical meaning. The coefficients become generic elasticities, not clearly associated with education or experience or nativity groups. And in the case of Borjas (2014), this generic elasticity is about 14, higher than the values reported in Borjas (2014), and implying more evenly spread impacts from immigration.

On the crucial question of immigrant-native substitutability, where does the truth lie? As usual we cannot be certain. The economic story we are interested in and competing stories we want to rule out can have effects on immigrant-native wage differences that are about the same within each education class and decade or each experience class and decade. *All* fall out of the analysis when BGH's extra dummies are added. If competing stories are at work, that will be enough for the regressions to bestow explanatory power on the dummies, arguing for their retention. What statistical tests cannot tell is whether the story of interest is also being expunged.

Since skill cell studies cannot assure us as to the truth, we must judge the probabilities as best we can by drawing in part on knowledge from outside these studies. Peri and Sparber (2009), reviewed just below, back the suggestion of imperfect substitutability by documenting that among low-skill workers, foreign-born ones work more in occupations such as construction that demand more physical skill and less communication skill. Friedberg (2001), one of the most credible studies in this review, finds that the massive influx of Russian Jews into Israel in the early 1990s boosted the earnings of those already in the Israeli workforce, again suggesting complementarity. And Cortes (2008), already reviewed, finds greater impacts on earlier immigrants.

Thus it seems likely that immigrants, especially recent ones, substitute imperfectly for natives. That may be good news for natives, but it is bad news for earlier immigrants.

## 3.4   Manacorda, Manning, and Wadsworth (2012), "The Impact of Immigration on the Structure of Wages: Theory and Evidence from Britain," *Journal of the European Economic Association*

Manacorda, Manning, and Wadsworth (2012) apply methods similar to OP's to data from the UK for the period from the mid-1970s to the mid-2000s. They find less substitutability than OP between immigrants and natives, which is to say, more complementarity between them. Their preferred estimate for the substitutability parameter is about 7.8, compared to OP's 20 (Manacorda, Manning, and Wadsworth 2012, p. 134). This suggests that in Britain the competitive effect of new arrivals concentrated even more on earlier immigrants, to the relative benefit of natives.

## 3.5   Summary: skill cell studies that do not exploit natural experiments

Skill cell studies arose in reaction to perceived limitations of spatial studies. Surprisingly, they seem to have done more damage to themselves than to the spatial literature, which has responded reasonably strongly to the methodological critiques levelled at them. If one trusts the theory-disciplined skill method, then the pattern of results in OP looks about right, with impacts concentrated on previous immigrants. If one joins

BGH in demanding much higher rigor, then about all we can say is that they produce no evidence that average effects are negative (for they cannot) and little evidence that the effects are concentrated among certain kinds of workers.

# 4  A study of immigrant-native complementarity
## Peri and Sparber (2009), "Task Specialization, Immigration, and Wages," *American Economic Journal: Applied Economics*

In reviewing the debate between Borjas, Peri, and their colleagues, I favored the Ottaviano and Peri (2012) view that natives and immigrants are probably not perfect substitutes, if only because natives' assimilation better suits them for some jobs. Peri and Sparber (2009) backs this common sense with systematic analysis. It is a spatial study with the unit of geography being the US state. It examines whether immigration leads natives to congregate more in occupations that demand more communication skill than physical skill, which would make sense since natives have a comparative advantage in communication in most parts of the country.

I have explained why spatial studies face challenges in discerning causality, and why the instruments typically used in them to improve matters, based on past migration patterns or (in this case) proximity to the US-Mexico border, are useful but not foolproof. Those concerns apply fully to the paper at hand. Indeed, contrary to a statement in the paper, econometric tests strongly challenge the validity of the instruments.[21]

Yet two features of the study lend it credibility. First, like Cortes (2008), its theory generates a *set* of intuitive predictions, all of which are borne out by the data, and which together seem best explained by the theory. In particular, states receiving more low-skill immigrant labor should see:

- For native and immigrant workers as a group, more labor in jobs demanding physical skill. After all, if the main effect of low-skill immigration is to expand the pool of workers suitable for such work, the economy should adjust to at least partially absorb them.

- But in response to the increased supply, wages for jobs demanding physical skill should fall relative to wages for jobs emphasizing communications skill.
- Following the money, and playing to their strengths, domestic workers should move more into communication-intensive professions. Thus the shift for this subgroup should be opposite that of the labor pool as a whole (see point 1).

The study's second strength is that, because the theory relates to how factors of production (different kinds of labor) compete in markets, it remains true even if causation runs in reverse, as will be explained just

---

[21] The instrumented regressions are overidentified, so the paper reports overidentification-based instrument validity tests and takes reassurance from them. However, the reasoning is problematic in several ways. First, inspection of the paper's code clarifies that the tests are only run for some of the regressions, e.g., in Table 2, only for those in the first row. Second, the test performed assumes homoskedasticity (Wooldridge 2002, p. 123), but the regression errors are clustered by state. Cluster-robust Hansen tests (Wooldridge 2002, p. 201) that I ran generally return worse results, most below $p = 0.1$. Finally, even if the reported test results are taken at face value, the fact that all those reported exceed 0.1 is less reassuring than the paper suggests. For example, pp values of 0.14 and 0.11 (Table 2, cols. 3 & 4) mean that if the instruments are valid, there is only a 14% or 11% chance of obtaining a test statistic as large as the ones actually obtained. If the purpose is to reach policy-relevant conclusions with confidence, those are low probabilities.

below.

All of the predicted patterns manifest strongly in the data (Peri and Sparber 2009, Tables 2 & 3). No doubt endogeneity is present even in the instrumented regressions (see my footnote 21). But that is less of a threat than usual because the theory is *reversible*. As in an animation of planets orbiting in reverse, the same laws of motion apply. If causation runs the other way too, it still indicates complementarity between immigrants and natives. The theoretical sequence would begin with disproportionate growth in industries demanding communication skills. This would raise wages for those skills relative to manual labor, and draw natives into those professions. This would secondarily increase demand for manual labor, which would attract immigrants with a comparative advantage in that work.

Given the strong match between the data and the subtle and complex predictions, and the consequent difficulty of explaining them with parsimonious alternative theories, the study lends credence to the view that natives and immigrants are not economically interchangeable, thus that the impacts of immigration concentrate to some degree on previous immigrants.

It should be noted that one trait that economically differentiates some migrants—being undocumented—is implicitly taken as fixed here but is in fact subject to policy. An expansion of legal immigration could affect not only the quantity of immigrants but the parameters that determine their impact on other workers. They might move more easily into customer-facing, communication-intensive professions.

# 5   Studies of wage and employment impacts exploiting natural experiments

The studies reviewed to this point look at correlations over years or decades between immigration, wages, and employment. Even those that instrument immigrant inflows are open to the standard criticism of endogeneity, that other dynamics explain the results found. For this reason, other studies exploit natural experiments: sudden, large influxes of workers. Even these are not beyond the standard doubts. E.g., the wave of Portuguese repatriates from Angola and Mozambique in the mid-1970s coincided with major economic and political changes in Portugal (Carrington and de Lima 1996). But natural experiment–based studies hold the promise of moving a step closer to the experimental ideal, and that promise should be investigated.

## 5.1   Card (1990), "The Impact of the Mariel Boatlift on the Miami Labor Market," *Industrial and Labor Relations Review*

Between April and October 1980 roughly 125,000 Cubans left home for Florida, on just about anything that could float. Perhaps half the immigrants settled in Miami, whose labor force they augmented by some 7% (Card 1990, p. 247–48). The Cuban government determined the start and the end of the episode. So the immigrants were not drawn out of Cuba by a sudden labor shortage in Miami—reverse causation that would have compromised the quality of the natural experiment.

Card (1990) examines the effects of this shock by comparing changes in wages and unemployment in Miami to those in Atlanta, Houston, Los Angeles, and Tampa–St. Petersburg, cities that received negligible numbers of Cubans in 1980. Card chose the comparator cities as similar to Miami in reliance on immigrants in the labor force, and in their patterns of economic development, as gauged by year-to-year employment growth circa 1980. The study fits into the spatial literature since it compares cities, while standing out in the suddenness of the migration inflow studied.

The conclusion is straightforward: the Mariel Boatlift had no discernible impact on wages and unemployment among those already in Miami. The main effect found is a drop in wages for Cubans in Miami—but that average *includes* the new arrivals and is about what would be expected given their number, relative youth, lack of education, and, thus, lower earning power (Card 2001, p. 254). Among blacks, a disproportionately low-skill group that might have been especially vulnerable to the extra competition for jobs, wages did not change relative to the control cities; and unemployment among blacks if anything improved in 1980. In 1982, unemployment among blacks did rise more in Miami than in the comparator cities, but given the timing that probably had much more to do with the dynamics of the deep 1981–82 recession.

Card (1990), p. 257, conjectures that Miami was unusually well prepared to absorb the influx because its economy already made disproportionate use of low-skill labor, thanks to earlier waves of immigration. Local industry such as textiles and apparel may have been able to expand quickly. That is a caution against generalization to the rest of the US or other countries.

Card (1990)'s most rigorous findings, such as those just cited on unemployment among blacks, are called "differences in differences." They are differences across cities in changes over time, which are themselves before-after differences. In a pedagogic article on methods in labor economics, the respected empiricists Joshua Angrist and Alan Krueger cite Card (1990) as an example of differences-in-differences (Angrist and Krueger 1999, Table 2). They also play off of Card's use of the method to illustrate its limitations:

> In the summer of 1994, tens of thousands of Cubans boarded boats destined for Miami in an attempt to emigrate to the United States in a second Mariel Boatlift that promised to be almost as large as the first one….Wishing to avoid the political fallout that accompanied the earlier boatlift, the Clinton Administration interceded and ordered the Navy to divert the would-be immigrants to a base in Guantanamo Bay. Only a small fraction of the Cuban emigres ever reached the shores of Miami. Hence, we call this event, "The Mariel Boatlift That Did not Happen." (Angrist and Krueger 1999, p. 1328)

Angrist and Krueger repeat some of Card's difference-in-difference analysis but centering on 1994 rather than 1980. Taking their results at face value, the "The Mariel Boatlift That Did not Happen" *increased* joblessness among blacks in Miami because their unemployment rate jumped from 10.1% to 15.1% between 1993 and 1994, even as the corresponding rate fell from 11.5% to 10.9% in the comparator cities. Of course, forces other than the immigration non-event must have caused that spike in joblessness.

Thus natural experiments are never perfect. The lesson for interpreting Card (1990) is that the actual Mariel boatlift *may* have raised unemployment among those already working in Miami, *if* this effect was coincidentally offset by unrelated economic developments.

Still, Occam's razor favors the simpler explanation, which is that immigration had little effect. Notably, Krueger later argued strongly for this view:

> The best available evidence does not support the view that large waves of immigrants in the past have had a detrimental effect on the labor market opportunities of natives, including the less skilled and minorities. Any claim that increased immigration…will necessarily reduce the wages of incumbent workers should be viewed as speculation with little solid research support. (Krueger 2006)

The first item in the "best available evidence" that Krueger invokes is Card (1990).[22]

## 5.2   Hunt (1992), "The Impact of the 1962 Repatriates from Algeria on the French Labor Market," *Industrial and Labor Relations Review*

When Algeria won independence from France in 1962, 900,000 ethnic French (*pieds-noirs*) repatriated to the motherland; their number was equivalent to 1.6% of France's labor force as of the next census in 1968. The repatriates resettled disproportionately in the south of France because its climate felt more like home (Hunt 1992, p. 556). The cross-province differences in resettlement rates created a natural experiment, which Hunt (1992) analyzes. Most of her follow-up data are from the 1968 census, so the impacts studied are considered medium- to long-term.

Overall, the study is less relevant to our interest than Card (1990) and less compelling. It is less relevant because the immigrants were more skilled on average than the receiving population (Hunt 1992, p. 557–58). No popular theory suggests that skilled immigration threatens the less-skilled who are our primary concern. These immigrants also probably assimilated more easily than most immigrants, meaning that they more perfectly substituted for, and competed with, native workers. That said, the inflow included low-skill workers too, so harm could be revealed for their native counterparts. And even a finding that the high-skilled immigration reduced employment or wages among already arrived counterparts would reveal something about the general economics of immigration.

At any rate, the study is less compelling because the "natural experiment" is less arbitrary. Since the immigrants favored the south and the impacts studied are longer-term, systematic north-south differences in economic development within France in the 1960s might explain the findings. "The general fall in unemployment from 1954 to 1962 and the rise in unemployment from 1962 to 1968 were particularly marked in the departments of the south." (Hunt 1992, p. 566.) Probably immigration was at most a secondary contributor to those trends. Indeed, provinces receiving more repatriates saw unemployment among non-repatriates rise more during 1962–68 (differences-in-differences regressions, Hunt (1992), Table 3, cols 3–4), which is what we would expect if disproportionate receipt of repatriates were coincidentally aligning with broader economic trends.[23]

Hunt finds large effects: the 1.6% labor expansion increased unemployment by up to 0.2 percentage points and cut wages by 0.51–0.80% (Hunt 1992, p. 566, 568). The upper bound of the wage impact implies a 2:1 ratio between labor supply growth and wage decline. But because the quality of the experiment seems low, I cannot confidently interpret these coefficients as causal.

## 5.3   Carrington and de Lima (1996), "The Impact of 1970s Repatriates from Africa on the Portuguese Labor Market," *Industrial and Labor Relations Review*

Carrington and de Lima (1996) is a close cousin of Hunt (1992) in title, journal, and source of identifying variation, looking at Portugal rather than France. However, the flow of *retornados* dwarfed the flow of *pieds-noirs* as a share of the receiving population. After Angola and Mozambique achieved independence in the mid-1970s, enough Portuguese repatriated to expand the country's labor force 10% in just three years. Like their French counterparts, they were better educated on average than the receiving population

Roodman, The domestic economic impacts of immigration

(Carrington and de Lima 1996, p. 335).

Despite the magnitude of the immigration shock, Carrington and de Lima express humility about their ability to trace its consequences:

> ...certain characteristics of Portugal in the 1970s make [the retornados immigration] a less-than-ideal natural experiment. First, Portuguese labor market statistics are certainly not as detailed and perhaps not as accurate as those of more developed countries such as France and the United States. More serious is the fact that the same years in which the immigration occurred brought the revolutionary end of a right-wing dictatorship in Portugal, the end of a colonial war, the nationalization of several industries, several labor market reforms, and a Europe-wide recession; it could therefore be that changes in the Portuguese economy over this period are not attributable to the retornados. (Carrington and de Lima 1996, p. 331)

Carrington and de Lima estimate the impacts in two ways: by comparing Portugal to France and, especially, Spain using national-level statistics; and by comparing across the departments within Portugal. In the latter case, data limitations force them to look only at wages in the construction industry.

Graphical comparisons show that Portugal paralleled France and Spain economically in the 1970s, with unemployment staying low until mid-decade, then climbing substantially. However, unemployment climbed sooner in Portugal, which might be a fingerprint of the *retornados.* (See Figure 1.) Corresponding time series regressions find that an additional immigration flow equal to 1% of the workforce was followed a year later by a 0.3% unemployment increase (Carrington and de Lima 1996, Table 4, col 2). These regressions control for the *Spanish* unemployment rate in an attempt to remove the effect of the mid-1970s European economic downturn. However, as with wages in Miami, the unemployment effects can be mathematically explained by high unemployment purely among the new arrivals. And separate regressions find that immigration was followed by a slight *increases* in earnings (Carrington and de Lima 1996, Table 4, col 6).

*FIGURE 1. UNEMPLOYMENT IN FRANCE, PORTUGAL, AND SPAIN (CARRINGTON AND DE LIMA 1996, FIG 3, PANEL*

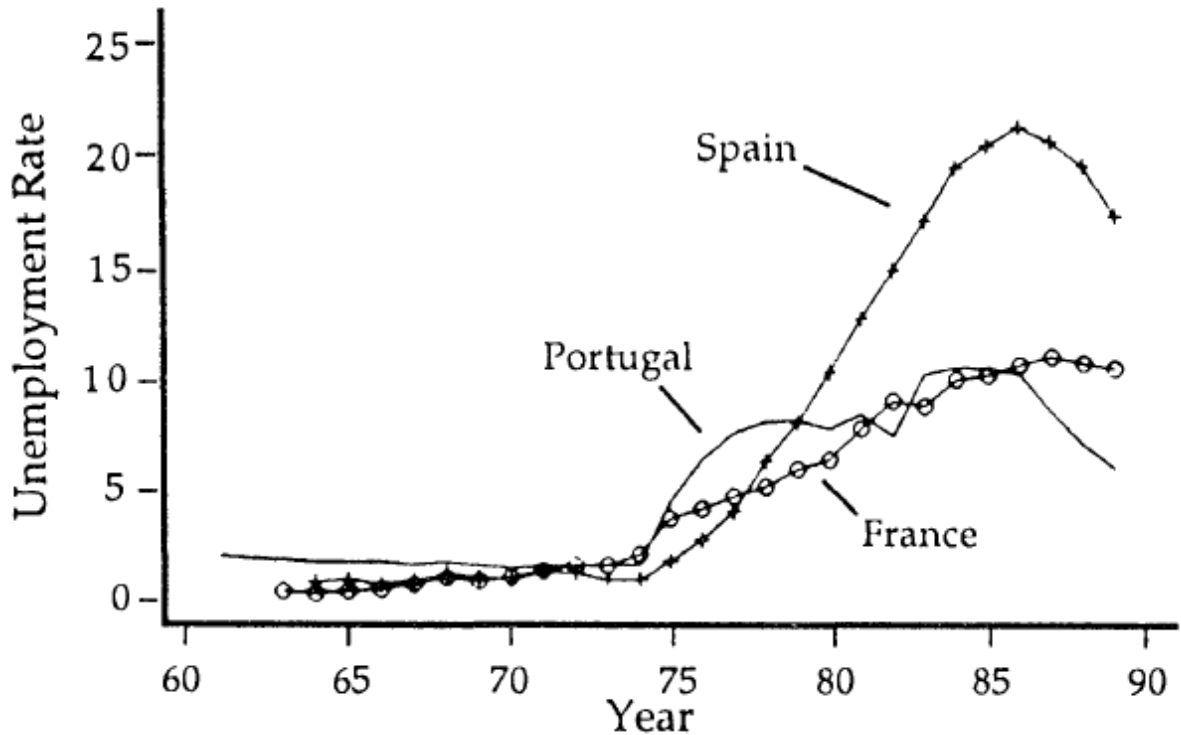

*A)*

Carrington and de Lima (1996) put even less weight on their within-Portugal, cross-department regressions. These show that wages rose less where population grew faster. However, Carrington and de Lima explain that the results appear driven by three outliers, all of which suffered economic hardships separate from immigration. Lisbon and nearby Setúbal "were the sites of the most dramatic political and economic conflicts during the mid-1970s. In addition, Setubal was the site of oil price–sensitive industries such as refining and ship-building" which were hit hard by the 1973–74 oil shock. Faro, at the country's coastal southern tip, suffered especially from the drop in tourism. Dropping these three, out of 18 departments, "shows a much weaker relationship between immigration and wages," although the authors do not report the precise effect. (Carrington and de Lima 1996, p. 343–44.)

Again, we find no strong suggestion of harmful side effects from immigration, which is perhaps unsurprising given the relative skill and ease of assimilation of the *retornados*.

## 5.4   Friedberg (2001), "The Impact of Mass Migration on the Israeli Labor Market," *Quarterly Journal of Economics*

Beating out Portugal in the World Cup of immigration shocks is Israel, which accepted Russian Jews equivalent in number to 12% of its population in the early 1990s. As in Portugal and France, but not Miami, the immigrants were relatively skilled, being both educated and experienced on average (Friedberg 2001, p. 1387). Unlike the studies of those events, Friedberg (2001) discerns impacts by looking at differences across occupations, not geographies. Did wages rise more slowly in the job types receiving the most newcomers?

That question is laden with potential for reverse causation since the immigrants did on average move into less-skilled, lower-paying jobs as they entered the Israeli economy at lower rungs. If occupations with lower pay also saw lower wage *growth*—if the wage premium for skilled professions widened in Israel in the 1990s—that would create the appearance of immigration slowing wage growth.

A strength of Friedberg (2001) is the use of a credible instrument for the share of workers in each occupation who were new immigrants, and it is made possible by Friedberg's decision to analyze across occupations rather than geographies. For each two-digit occupation code and year, the instrument is the number of émigrés who worked in that occupation *before* they left Russia, taken as a share of Israel's pre-inflow stock of workers in that occupation. Since both the top and bottom of that ratio were established before the great migration, the migration could not have reversed-caused the instrument. And it is hard to think of third variables that could have influenced the occupational distribution of Jews in Russia in the 1980s and the evolution of wages by occupation in Israel in the 1990s.

Friedberg's results are striking. A 12% workforce expansion from immigration gave those already in the workforce an 8.9% raise on average over 1989–94 (Friedberg 2001, p. 1395, after scaling from a hypothetical 10% inflow to the actual 12% one). With much less statistical precision, she also finds that Russian entrants into an occupation increased the number of natives in the occupation, in a 1:2 ratio. These patterns fit Ottavani and Peri (2012)'s model of complementarity between foreign- and native-born workers.

Friedberg cites case studies that lend granularity to these patterns:

> *Sussman and Zakai [1998] study the labor market for physicians in Israel in the early 1990s. They find that Russian physicians—even those with considerable prior experience and expertise—were confined to positions as generalists at the lower end of the pay scale in Israeli hospitals. This enabled native Israeli physicians to be promoted to fill the higher-paying ranks in an expanding health-care system. In addition, because the Russians relieved the overall staffing burden at hospitals, native Israelis were able to devote more time to yet higher-paying private-practice work. Sussman and Zakai conclude that "relations between the two groups were complementary rather than competitive…in providing medical services."*

> *In the case of engineers and other highly educated specialists, a government committee [Eckstein et al. 1996] gathered evidence on the nature of their employment. Firm managers testified that Russian workers were often assigned to more basic tasks or supportive roles, freeing Israelis to work on the more productive aspects of projects. As in the case of physicians, Russian workers filled positions at the lower end of the job ladder, pushing incumbent Israelis up the ranks into more supervisory, high-paying roles. (Friedberg 2001, p. 1396)*

Friedberg (2001) makes a compelling case that the Russian influx into Israel benefitted those already there, in part because of the challenges of assimilation, which consigned the immigrants to junior work roles despite their depth of education and experience. Her study suggests that industrial economies have the flexibility to absorb new workers quickly.

One important caveat for our review is that the story might have been different for a predominantly low-skill immigrant cohort. Also, Israel constructed institutions to facilitate assimilation—not least, automatic citizenship for Jews—that will be absent in other contexts.

## 5.5   Clemens (2013a), "The Effect of Foreign Labor on Native Employment: A Job-Specific Approach and Application to North Carolina Farms," working paper

Clemens (2013a) is an original paper with compelling econometrics. It appears unique in studying changes not in the supply of immigrant workers who might compete for certain jobs, but in the demand among natives for those jobs. And it exploits a sharp natural experiment: the Great Recession, which caused the number of Americans looking for work to spike.

Clemens studies whether higher unemployment in North Carolina after 2008 led more people to take farm labor jobs that would otherwise have gone to migrant workers with H-2A visas. Under the H-2A program, employers wanting to hire temporary immigrants must first advertise the positions in local media, give priority to American applicants, and report to the government on these activities, giving justification if any American applicants are not hired (Clemens 2013a, p. 11–12). The intent is to ensure that the visas deprive no American of a job. Regardless of who fills the positions, they are required to pay at least the minimum wage.[24]

As Clemens emphasizes, these legal requirements add a second distinctive dimension to the natural experiment: in principle, employers must manifest the same economic demand for native and foreign workers. In principle, then, any differences in the rate at which the two groups take the work must derive from differences in their willingness to do so rather than in their value to employers—that is, from the supply side of the labor market, not the demand side. If natives avoid the work, "this might be because natives dislike manual or routine work itself, because they dislike circumstances of the work (dirt, stench, exposure to the elements), or because they incur a social stigma for performing such work." This is scientifically interesting because conventional studies—including all above—cannot distinguish supply- from demand-side factors.

Clemens's result can be stripped down to a few stark numbers. After the unemployed population in the North Carolina counties studied surged from 283,000 in 2008 to 490,000 in 2009, the number of natives who sought such agricultural work (in the sense of getting a referral from an unemployment agency) *fell*, from 170 to 108. The number who actually started the work dropped from 58 to 48. And the number who stuck it out for the season fell from 11 to 6. (Clemens 2013a, Table 2.) In sum, more than 200,000 people lost their jobs and essentially none went to work picking produce. This is convincing evidence that at today's margins, Americans and migrants are not competing for these jobs. An expansion of migrant farm worker programs would therefore cost essentially no American jobs.

Like the other natural experiment–based studies, this one gains internal validity (convincing econometrics) at the expense of external validity (generalizability). Admitting more Jamaicans temporarily to work fields in North Carolina is a different thing from admitting new, low-skill immigrants as permanent residents and citizens. The two may affect labor markets differently. In this case, in thinking about generalizability, I think it helps to understand better the economic disconnect between temporary migrant farm workers and their erstwhile American competitors.

Barriers to immigration between rich and poor countries are like dikes. Even if they leak a bit they sustain differences between the water levels—wages—on each side. As noted at the outset of this review, Clemens, Montenegro, and Pritchett (2008) have documented just how well the dikes work: "observably identical" people earn 2.0–15.5 as much on the rich side as on the poor. The thrust of US migrant farm worker programs is to admit foreigners onto rich-country soil while keeping them on the other side of the dike legally, politically and economically. Jamaican workers may come to America but the employers come to the Jamaican labor market. If American law requires employers to treat Jamaicans like Americans, to pay Jamaicans more than would be needed to attract them to the jobs, then we should expect employers to tacitly claw back some of those "rents" (i.e., the windfall to Jamaicans from being paid like Americans). As profit maximizers, agricultural firms will likely aim to reduce the effective cost of the immigrants and seek to hire just them. They might find ways to pay the immigrants less, such as by paying for piecework that must be done at unrealistic rates to match minimum wage. And they might cut corners on housing, food, and worker safety (such as protection from pesticides) while leveraging their market power to punish those who complain about conditions. They might buy support in Congress and capture the bureaucracy that regulates them. They might try to shed native workers, who may stir up trouble. All these things have been documented to some degree (Yeoman 2001, Guerra 2004, Farmworker Justice 2011, Jornaleros Safe 2013). The deep force at work is that the immigrants, as foreigners, have less recourse than native citizens and so in equilibrium such leveraging of power takes place more than if the workers were native.

If right, this insight has interesting implications for the generalizability of the Clemens (2013a) finding. On the one hand, it explains why a non-marginal change—abolition of migrant farm workers—*would* benefit native low-skill workers in the first instance. To entice them, farm managers would have to offer better wages and/or working conditions. (Though the workers might be hurt indirectly by higher food prices and lower local spending by skilled employees in the weakened local agriculture industry.) This scenario, though, is academic, since it is not a serious political possibility. What is more realistic is expansion of migrant farm labor, to which the Clemens (2013a) finding generalizes.

Meanwhile, the finding of low native-immigrant competition at the current margin may generalize in spirit to a huge class of people: undocumented migrants, including a lot of those kitchen workers. They too lack the economic and legal recourse of natives, which weakens their bargaining power in the labor market. Perhaps this helps explain some of their apparent complementary with natives. Perhaps if they gained legal status en masse they would compete more directly with all legal workers, diffusing their competitive wage effects more widely.

## 5.6 Summary: studies of wage and employment impacts exploiting natural experiments

Unsurprisingly, the studies based on natural experiments produce some of the most compelling evidence in this review. Particularly strong are Card (1990) on the Mariel Boatlift, Friedberg (2001) on Russian Jews entering Israel, and Clemens (2013a) on migrant farm workers in North Carolina. Subject to important caveats about generalizability—e.g., can all cities absorb Cubans as fast as Miami or Israel?—they provide strong evidence of zero or positive side effects for natives.

## 6 Studies of impacts on innovation and productivity

The studies reviewed so far have mainly looked at impacts on earnings and employment. A distinct but related economic variable potentially affected by immigration is *productivity*, which is ultimately driven by *innovation*. The observation that motivates an interest in these variables is that skilled foreign workers are extremely important in technology industries, which is why big software companies lobby for more openness to immigration. The prominence of immigrants in the skilled workforces of innovative companies suggests that immigration can affect the pace of innovation, with long-term implications for all workers.

The study of productivity effects is also an important complement, as it were, to skill studies, which, recall, can only assess relative, not absolute, effects on wages. If we believe that immigration raises productivity and thus wages for large swaths of workers, then a skill-cell finding that some groups' wages fall *relative* to others' might only mean that wages rise more slowly for some than others.

## 6.1 Kerr and Lincoln (2010), "The Supply Side of Innovation: H-1B Visa Reforms and US Ethnic Invention," *Journal of Labor Economics*

The reviews above of the two Peri, Shih, and Sparber studies describe the US H-1B visa program, which licenses employers to import skilled labor. Kerr and Lincoln (2010) also study the impact of the H-1B program using spatial methods, but their primary interest is not in wages and employment, but in patenting as an indicator of innovation.

Since patent records do not reveal the immigration status of the filers, Kerr and Lincoln track whether the number of filings from a given metropolitan area by people with Indian or Chinese names varied in tandem with the estimated local stock of H-1B workers of the same nationality. About 40% of H-1B recipients during 2000–05 came from India, and another 10% from China.

Despite the standard concerns about spatial studies, the *pattern* of results once again helps convince, by tending to rule out alternative theories. During 1995–2007, among cities in the top quintile of dependence on H-1B workers, a 10% increase in the national H-1B population corresponded with a 6–12% increase in patent filing by people with Indian or Chinese names and a 0–2% rise in filings overall (Kerr and Lincoln 2010, p. 498). The effects are smaller for lower-H-1B-dependence quintiles, and for nationalities less prominent in the H-1B visa program (Kerr and Lincoln 2010, Table 4, col 3 and rows 2& 3). This is what one would expect if immigration is indeed causing more patenting. There was no effect on patenting by people with Anglo-Saxon names—again as one would expect, since the population of such people would hardly have grown from H-1Bs (Kerr and Lincoln 2010, Table 4, col 4). And in an "placebo" test, variation in H-1B workers in the US had no effect on US patent filing rates by people in Canadian cities (Kerr and Lincoln 2010, Table 4, panel E). That might seem like a strange test, but to the extent that the US and Canadian technology industries evolve similarly, it helps rule out the possibility that something about the American technology industry was driving both higher patenting and more recruitment of foreign workers, without causation running from immigration to invention. For if it had, Canadian patenting would probably have risen too.

Overall, the study supports the view that immigrants contribute to American innovation. It finds little sign that they displace natives as inventors.

## 6.2 Peri (2012), "The Effect of Immigration on Productivity: Evidence from U.S. States," *Review of Economics and Statistics*

This paper shares features with Peri and Sparber (2009) and Peri, Shih, and Sparber (2014a), reviewed above, while studying impacts on productivity rather than job specialization or wages. It is spatial, with the unit of geography being the state; its instruments are similar, being based on both past immigration patterns and proximity to immigration ports of entry; it includes robustness tests that attempt to address challenges to the instruments; and yet, as in Peri and Sparber (2009), the instruments do worse on econometric tests of validity than the paper's contends.[25] Unfortunately the paper's conclusions are weaker than those in Peri and Sparber (2009) because the statistical findings are consistent with plausible competing theories. The positive correlation discovered between immigration and state-level productivity growth can easily be explained by higher productivity growth leading to higher economic growth, leading in turn to more work opportunities for migrants. The failure of the instruments on statistical validity tests leaves the door open to this explanation.

## 6.3   Borjas and Doran (2012), "The Collapse of the Soviet Union and the Productivity of American Mathematicians," *Quarterly Journal of Economics*

The collapse of the Soviet Union made several demographic waves—not just the flow of Jews to Israel. Among the tiniest was the emigration of some 336 Soviet mathematicians to the United States, where many took up residence at elite universities in the early 1990s ((Borjas and Doran 2012, p. 1161).

After voluminous data collection and painstaking analysis, Borjas and Doran (2012) show convincingly that the intensified competition from immigrants reduced job opportunities for American mathematicians, forcing some into other professions. Productivity of American mathematicians, measured by papers published, also fell. The study thus supports the Ottaviano and Peri (2012) finding that natives and immigrants compete more directly at the high end of the skill spectrum, while, ironically, tending to contradict the Friedberg (2001) study of Russian Jewish immigration to Israel.

However, the study seems among the least generalizable in this review. The elite university operates on a peculiar business model. To recruit top professors, the universities offer a rent: substantial time to perform research as the professors choose, which does not really create capturable economic value for the education industry, but strengthens the university in the zero-sum competition for students and funding. The amount of this rent available nationally grows only slowly, so it makes sense that the arrival of great mathematicians from the Soviet Union displaced less-talented Americans. But the rest of the private sector probably has more capacity to absorb talent in ways that create fresh economic value, which employers can partly capture. Outside of the academy, this should generally make hiring additional talent a positive-sum game and remove the sort of ceiling found for academic mathematicians.

## 6.4   Moser, Voena, and Waldinger (2014), "German Jewish Émigrés and U.S. Invention," *American Economic Review*

This new study arrives in counterpoint to Borjas and Doran (2012). Moser, Voena, and Waldinger (2013) estimate the impacts on American innovation of the migration of Jewish chemists from Nazi Germany to the United States.[26] Why chemistry? Through to the early 20th century, Germany led the discipline, so many of those who fled were among its giants. After moving, the émigrés revolutionized American chemistry (Moser, Voena, and Waldinger 2013, p. 3222). Meanwhile, innovation in chemistry is easier to patent than in, say, math and physics, so it leaves clearer footprints for economists to follow, in the form of patent filings.

Moser, Voena, and Waldinger make comparisons across subfields of chemistry, according to a 166-entry typology, asking whether those receiving any of the 26 émigrés to the US saw more patenting by American nationals.

---

[25] The comments in footnote 21 apply here. As an example, in my runs, the total factor productivity regressions (Peri 2012, Table 2, row 4) return Hansen overidentification test *p* values of 0.01, 0.11, 0.15, 0.27, and 0.15, with smaller values appearing alongside larger estimates of the impact on productivity.
[26] The study also covers chemists from Austria, which Germany annexed in 1938.

One endogenous-causation story to worry about is that US patent activity *followed* German innovation for reasons aside from immigration. Perhaps a decade or so after Germans made breakthroughs in basic science, Americans turned those insights into patentable inventions. We might then expect patenting in the subfields receiving German émigrés to take off simply because those also tended to be the subfields with recently active German researchers, émigré or otherwise. To guard against this possibility, Moser, Voena, and Waldinger restrict their sample to subfields with active German researchers—émigré or otherwise— where "active" is defined by filing for US patents. (Even then, foreign nationals could and did file for US patents.) Within this subset, they perform differences-in-differences, examining whether patenting by Americans rose more in subfields not only having active German researchers, but receiving German émigrés, as compared to fields with active German researchers but no émigrés.

To further reduce endogeneity, the authors borrow a trick from the Friedberg (2001) study of Russian Jews in Israel. Since émigrés might have switched fields after arrival in the US to work in areas where Americans were more innovative anyway—another endogeneity story to rule out—Moser, Voena, and Waldinger create an instrument that assigns émigrés to their pre-arrival subfields. In effect, they search for a patenting bump-up among Americans in subfields in which German Jews were active before the migrated, in comparison with subfields in which only non-migrating Germans worked.

This design makes for persuasive conclusions. A representative result: Chemistry subfields with émigrés recorded an extra 170 patents/year during 1933–70, equivalent to an increase of about 70% over pre-1933 levels (Moser, Voena, and Waldinger 2013, p. 3239).

Digging deeper, the authors find that 75% of the relative increase was associated not with established US inventors, but with ones who had not patented before 1933 and were probably younger and newer to chemistry (Moser, Voena, and Waldinger 2013, p. 3247). Cutting the data another way, the increase occurred more among US scientists who were co-inventors of the émigrés, meaning ones whose names appeared alongside an émigré's in at least one filing—and among co-inventors of co-inventors. These findings combine to suggest that émigrés shaped American chemistry through network effects, especially by attracting students to their subfields. In this way, immigrants complemented natives even as they must have competed with them too.

The positive side effects of German immigrants for native chemists contrasts with the Borjas and Doran (2012) results on Soviets in mathematics. What to make of the contrast? Both studies are credible, so the relevant question is not which is right, but which is more representative of the rest of the economy. Notably, while patents are only a rough proxy for the economic value of innovation—many great ideas, such as the World Wide Web, are not patented while many insignificant inventions ones are—patent counts are probably more meaningful than paper publication counts, the key productivity indicator in Borjas and Doran (2012). Patents are closer to the economic process of innovation and the capacity of the US Patent Office to absorb innovations is not capped like journal space. Since the capacity of an industrial economy to exploit innovation is also not limited by quota, the impact of immigration on patent filings in chemistry looks to be more representative of the typical impacts of skilled immigration on productivity.

On the other hand, the typical migrant is not a titan of research, like some among those 26 émigrés. So we should not rush to cite this study as proof that skilled immigration speeds innovation among natives generally.

## 6.5   Summary: studies of impacts on innovation and productivity

This mini-review of research on how immigration affects innovation and productivity is motivated by two sorts of implications. First the research speaks to whether immigration of skilled professionals likely helps the majority of the receiving population in the long term by accelerating innovation, in a way that is hard to pick up in direct analysis of impacts on wages and employment. Second, the research can provide more examples of immigrants affecting labor markets, which is especially interesting to the extent that we can generalize from the cases studied to the low-skill workers we care most about because of their vulnerability.

It bears emphasizing that under neither motivation should we shed many tears for any skilled natives who lose jobs or pay from immigration. Perhaps some of those displaced mathematicians ended up working on Wall Street. As Clemens (2011), p. 95, points out, the side effects of immigration, positive and negative, arise out of the operation of markets. This, economists normally do not regret because it increases efficiency. The main economic caveat comes if efficiency exacts a cost among people who are poor.

With regard to the first motive, only one of the studies reviewed in this section attempts to link immigration directly to productivity (Peri 2012), and it is open to substantial challenges. So there is little strong evidence that immigration broadly raises economic productivity and thereby helps most workers in a way not picked up in research that looks more directly at wages. The three studies of impacts on innovation, proxied by journal papers and patent filings, are more convincing. Whether those patent filings translate into higher productivity broadly shared again seems conjectural.

As for the second motive, the study showing negative side effects (Borjas and Doran 2012) seems least generalizable, especially to markets for low-skill labor. And Kerr and Lincoln (2010) and Moser, Voena, and Waldinger (2013) find neutral-to-positive impacts on natives.

## 7   Conclusion

Table 2 summarizes this review.

The evidence base does not support us with certainties, only best bets. In this case, a variety of studies deploying different methods in different contexts, and some general knowledge about how economies work, coalesce into a fairly consistent picture. Industrial economies can generally absorb migrants fairly quickly, in part because capital is flexible, in part because immigrants are consumers as well as producers. Thus, long-term average impacts are probably zero at worst. And the long-term may come quickly, especially if migration inflows are predictable to investors. Possibly, skilled immigration boosts productivity and wages for many others, but at this point that is mostly a matter of conjecture.

Around the averages, there are distributional effects. The labor market does appear somewhat segmented, not just by skill level, as measured by education and experience, but also by nativity, which itself may proxy for possession of skills such as fluency in the receiving country's language. Thus the competitive side effects of immigration to a degree concentrated within certain labor market segments. Among low-income workers, the ones who stand to lose the most are those who most closely resemble new arrivals, in being immigrants themselves, being low-skill, being less assimilated, and perhaps in being undocumented. Thus the common statement that *natives* have little to fear does not represent the whole story.

By the same token, native workers and earlier immigrants tend to *benefit* from the arrival of workers different from them, who complement more than compete with them in production. An important consequence is that skilled immigration can offset the effects of low-skill immigration on natives and earlier immigrants.

The third and fourth columns of Table 1, based on Ottaviano and Peri (2012), display some representative estimates of the impacts of the large and relatively skill-balanced US immigration inflow of 1990–2006. A roughly 100% increase in the immigrant stock is estimated to have raised wages slightly for natives while reducing them about 10% among less-educated earlier immigrants. Arguably these losses are modest compared with the earnings gains the earlier immigrants themselves achieved by coming to the US.

Looking ahead, a proportionally larger or less skill-balanced immigration inflow could have larger impacts. A more *legal* inflow, on the other hand, might reduce the economic ghettoization of immigrant low-skill workers and diffuse their competitive impacts more evenly.

*TABLE 2. SUMMARY OF STUDY REVIEWS*

| Study | Unit of observation | Source of variation for cited findings | Findings | Comments |
|-------|---------------------|----------------------------------------|----------|----------|
| *Spatial studies that do not exploit natural experiments* | | | | |
| Card (2001) | US city × broad skill/occupation group, 1990 | Estimates of 1985–90 immigration to city-occupation groups, based on total US immigration by sending country, their national-level dispersion across education groups, and 1985 immigrant stock by city and sending country. | 10% labor supply increase cuts employment rate 2.02 points (male natives), 0.81 points (female natives), 0.96 points (male earlier immigrants), 1.46 points (female earlier immigrants); cuts wages 0.99 points (male natives), 2.5 points (female earlier immigrants), indistinguishable from 0 (other groups). | Blunts criticisms of spatial studies by focusing on short run and showing that immigration not followed by departure of natives from a city. Looking across cities and skill-levels at once allows removal of bias from omitted third variables with impacts that are constant across either dimension, such as some cities having more dynamic economies. But impacts on earlier immigrants only indirectly inferred, from employment and wage changes in occupation groups they tend to enter. |
| Cortes (2008) | US city × census round, 1980–2000 | Estimated low-skill immigrant stock by city, year based on national-level stocks by sending country and 1970 stocks by city and sending country. | 10% increase in the immigrant share in low-skill workforce cuts prices of low-skill–intensive services 2% and low-skill wages ~1–1.5%. | Weak-to-no impact on high-skill–intensive services and tradable goods corroborates interpretation of correlation as causation since full pattern hard to explain otherwise. |

| Study | Unit of observation | Source of variation for cited findings | Findings | Comments |
|-------|---------------------|----------------------------------------|----------|----------|
| Peri, Shih, and Sparber (2014a) | US city × period (periods: 1990–2000, 2000–05, 2005–10) | Estimated H-1B–driven rise in STEM workforce, based on 1990 foreign STEM workforce by city and sending country and national-level distribution of H-1B visas by sending country | A 1 percentage point increase in foreign share in STEM workers raises native STEM wages 7–8% | Impacts implausibly large: can explain 30–50% of US productivity growth in 1990–2010. Suggests endogeneity. |
| Peri, Shih, and Sparber (2014b) | US cities, 2009–11 | Numbers of H-1B visa denials for computer-related professions, which depend on numbers of applications and lottery | 177,386 visa denials in 2007–08 cost 231,000 native computer-related jobs | Reliance on lottery as random determinant unpersuasive because it likely mattered less for number of visa denials than did (non-random) number of applications. |

***Skill cell studies that do not exploit natural experiments***

| Study | Unit of observation | Source of variation for cited findings | Findings | Comments |
|-------|---------------------|----------------------------------------|----------|----------|
| Borjas (2003) | Education level × experience level × US census survey, 1960–2000 | Number of foreign-born workers in each education-experience-year group | ~10% migration-induced labor growth in 1980–2000 cut wages for native non-high-school-completers 8.9% | Assumes the capital stock did not adjust to labor expansion and that native- and foreign-born are perfect substitutes so that natives face as much competition from new arrivals as do earlier immigrants. |
| Borjas (2014), ch. 5 | Education level × experience level × US census survey, 1960–2010 | Number of foreign-born workers in each education-experience-year group | 10.6% migration-induced labor growth in 1990–2010 cut wages for native non-high-school-completers 6.2% (no capital adjustment) or 3.1% (after full capital adjustment) | Argues that native- and foreign-born are best seen as perfect substitutes, as above. |
| Ottaviano and Peri (2012) | Education level × experience level × native/immigrant × US census survey, 1960–2006 | Total person-weeks of work in each education-experience-nativity-year group | 10.0% migration-induced labor growth in 1990–2006 *raised* wages for native non-high-school-completers 1.7%, cut them for foreign-born 8.1% (both after full capital adjustment) | Finding of imperfect native-immigrant substitutability differences more fragile than Borjas findings on substitutability across skill groups, as shown in Borjas, Gorgger, and Hanson 2012. |
| Manacorda, Manning, and Wadsworth (2012) | Education level × experience level × native/immigrant × UK survey year, ~1975–2005 (quinquennial) | Number of people in each education-experience-nativity-year group | Migration-induced labor growth in 1975–2005 reduced wages for immigrant university graduates 0.8%/year compounding; no effect on others | |

***Studies of wage and employment impacts exploiting natural experiments***

Roodman, The domestic economic impacts of immigration

| Study | Unit of observation | Source of variation for cited findings | Findings | Comments |
|---|---|---|---|---|
| Card (1990) | Miami + 4 other US cities, early '80s | Occurrence of Mariel Boatlift from Cuba in April–October 1980, which expanded Miami's workforce 7% but the other cities' hardly at all | No change in following few years in wages or unemployment in Miami relative to comparator cities—notably among blacks, a disproportionately low-skill group | Natural experiment not as perfect as randomized since other events could have offset Boatlift's effects. But zero change after 7% labor supply spike most easily explained as (non-)impact. |
| Hunt (1992) | Provinces of France, 1967/68 | When Algeria won independence in 1962, 900,000 ethnic French, relatively skilled on average, repatriated, settling in some provinces more than others | Nationally, 1.6% labor supply growth increased unemployment by up to 0.2 percentage points, cut wages 0.51–0.80% | Experiment may not have been clean since repatriates settled more in the south of France, whose economic trajectory differed in the 1960s. |
| Carrington and de Lima (1996) | Portugal vs. Spain, 1970s; districts of Portugal, 1970s | Repatriation of ethnic Portuguese from Angola and Mozambique, relatively skilled on average, in 1974–76 | Sudden 10% labor expansion may have sped Portuguese unemployment rise during European downturn and depressed construction wages | Experiment may not have been clean because of simultaneous major events in the country: revolutionary end of a dictatorship, nationalization of industries, labor law changes. |
| Friedberg (2001) | Large sample of Israeli workers, divided into some 100 occupation groups, 1989 & 1994 | Flow of Soviet Jewish immigrants to Israel in 1989–95, by pre-departure occupation | Sudden 12% labor expansion *increased* earnings 8.9% for those already in workforce over 1989–94 | Persuasive. Hard to otherwise explain correlation between number of immigrants by pre-departure occupation and later Israeli wage growth in same. Backed by case studies. Suggests complementarity between new immigrants and rest, at least among skilled workers. But Israeli efforts to absorb migrants, including instant citizenship, perhaps rare. |
| Clemens (2013a) | 58 employment offices × 66 months, North Carolina, Feb 2005–May 2011 | Great Recession–caused unemployment jump 2008–09 | After total unemployed in studied counties rose from 283,000 to 490,000, number of natives who took and held farm jobs normally filled by migrants went from 11 to 6. | Persuasive. Under current pay and working conditions, natives avoid farm labor. Migrant farm labor market separate from native job market. |

| Study | Unit of observation | Source of variation for cited findings | Findings | Comments |
|---|---|---|---|---|
| **Studies of impacts on innovation and productivity** | | | | |
| Kerr and Lincoln (2010) | US cities × year, 1995–2007 | Estimated number of H-1B holders in a city, by ethnicity, based on national-level H-1B ethnic breakdown and number of H-1B applications in 2001–02 | Among top quintile of cities in H-1B dependence, 10% increase in national H-1B population associated in same year with 6–12% increase in patent filing by people with Indian or Chinese names and 0–2% rise overall | Persuasive. Chinese and Indian largest ethnicities among H-1B holders. Small-to-no effect found for less-common H-1B ethnicities, for filers with Anglo-Saxon names, for less H-1B–dependent cities, and for Canadian cities (placebo) test. Pattern most easily explained as impact. No sign of displacement of native inventors. |
| Peri (2012) | US states × US census survey, 1960–2006 | Distance to Mexican border; estimates of migrant stocks based on 1960 stocks by state and sending country, and national-level growth rates by sending country since. | Immigration increases productivity (output per units of labor and capital input) | Study performs aggressive robustness tests. However, contrary to study text, instruments fail an exogeneity test and results can be explained by non-exogenous theory such as reverse causation. |
| Borjas and Doran (2012) | US & Soviet mathematicians who published in 1970–89 | Arrival of ~336 Soviet mathematicians in US just after collapse of USSR | American mathematicians in subfields with active émigrés were published and cited less after 1992 and more likely to leave profession, indicating zero-sum displacement by new immigrants. | Persuasive. But does not seem a good model for most of the economy. Academic postings and journal space grow slowly, so that new talent, if superior, must displace old. Scope for innovation in most of the economy is more elastic. |
| Moser, Voena, and Waldinger (2013) | 166 chemistry subfields × year, US, 1920–70 | Starting in 1933, arrival of 26 Jewish émigré chemists from Nazi Germany & Austria, distinguishing their pre-departure subfields from ones with active German/Austrian researchers who didn't leave | American inventors in subfields with émigrés recorded an extra 170 patents/year in 1933–70 in total, 70% over pre-1933 level | Persuasive. As in Friedberg (2001), correlation with émigré's pre-departure field reduces plausibility of competing theories. Backed by historians' observations of the impact of the émigrés on chemistry in US. |

## Appendix

Here I discuss the criticisms of Ottaviano and Peri (2012) (OP) in Borjas, Grogger, and Hanson (2012) (BGH).

### Errors in data set construction

In replicating the construction of the OP data set, BGH find errors, such as failing to exclude self-employed workers when computing average wages, which is not what OP intended. In my experience, replication always turns up such errors.[1] BGH show these fixes to increase the apparent substitutability between immigrants and natives. In the empirics cited below, I use Borjas (2014) data, which should incorporate these corrections.

### The fragility of estimates of the substitutability of high school graduates and high school non-completers

BGH argue that just as the relationship of labor and capital resists estimation in national data, so does the substitutability between workers with and without a high school degree. But BGH do not tie this point directly to the OP analysis and results. Rather, they take it to "illustrate the limits of what the nested CES framework [introduced by Borjas (2003)] can teach us about the labor market impact of immigration." The argument thus does not seem to speak to the relative credibility of OP and the Borjas analyses but to the value of the whole skill cell approach. Also, as argued below, the ordering of the CES hierarchy is not sacred. The split in question can be pushed lower, providing a larger sample for estimation.

### Log mean wages versus mean log wages

In aggregating census data for individual workers to the level of the skill cell, OP take the log of the ratio of the averages of immigrant and native wages, which is the difference of the log means. BGH contend that it is the *mean log* that is theoretically valid and standard practice. And it happens to matter: in BGH's replication, making the switch destroys OP's finding that natives and immigrants are somewhat complementary.

Using mean logs does seem to be standard practice, although one precedent BGH cite, Katz and Murphy (1992), Table 1, appears to take log means. Yet at least in the context at hand, theory seems to side with OP. This does not vitiate the concern about fragility in the OP regressions—results should be robust to this sort of mathematical niggling. But it does mean that the criticism is a bit overdrawn. The BGH versions of the OP regressions are not obviously superior.

The theoretical question is: what production function best describes how individual workers contribute to total productivity-adjusted labor supply in a skill cell? The most natural assumption is that workers within a cell are perfectly substitutable, so that their total labor is a linear combination of their individual contributions. As a simple example to get at functional form, suppose the economy is Cobb-Douglas with

---

[1] Indeed, in working with the Borjas 2014 code for his Table 5.3, I found an error in the regressions estimating native-immigrant substitutability for male and female workers together: they use male-only wage variables for immigrants. Fixing the error does not change interpretations.

Roodman, The domestic economic impacts of immigration

inputs capital, $K$, and perfectly substitutable workers, $L_i$:

$$Q = L^\alpha K^{1-\alpha} = \left( \sum_i \theta_i L_i \right)^\alpha K^{1-\alpha}$$

The wage of worker $i$ is taken to be her marginal product, which is

$$w_i = \frac{\partial Q}{\partial L_i}$$

So the mean wage, weighted by hours worked, is

$$\bar{w} = \frac{\sum_i L_i w_i}{\sum_i L_i}$$

How are we to meaningfully define a marginal increase in aggregate labor $L$ in order to speak of the marginal product of aggregate labor? The best way seems to be to equate a marginal increase in $L$ to simultaneous marginal increases in each $L_i$ in proportion to each worker's share in total hours, i.e., in proportion to $L_i$. So the marginal product of aggregate labor would be defined:

$$\frac{\partial Q}{\partial L} \equiv \frac{\sum_i L_i \frac{\partial Q}{\partial L_i}}{\sum_i L_i} = \frac{\sum_i L_i w_i}{\sum_i L_i} = \bar{w}$$

and

$$\log \frac{\partial Q}{\partial L} = \log \bar{w}$$

Thus the log mean used by OP is a theoretically grounded proxy for the log marginal product of labor.

## "Correct" weights?

BGH argue that if we grant OP their log means, OP still use the wrong observation weights. The usual rationale for weighting in this literature is to increase efficiency (precision) by down-weighting observations that are less precise, coming from smaller underlying samples.

Since the OP dependent variables are logs of ratios of mean wages, BGH write down the formula for the variance of such variables, whose inverse is the "correct" weight. The BGH weights are sensible; but calling them "correct" again overreaches. This is because in the heteroskedastic regression $y_i = \beta x_i + \epsilon_i$ with $\text{Var}[\epsilon_i] = \sigma_i^2$, the efficient weight is not $\text{Var}[y_i]$, which BGH compute, but $\text{Var}[y_i|x] = \text{Var}[\epsilon_i] = \sigma_i^2$, which is unknown. Since we are ignorant of the efficient weights, we cannot be certain that the BGH weights are closer to them than the OP ones.

## Saturation with dummies

BGH point out that the addition of a large number of fixed effect dummies destroys the OP estimates of native-immigrant substitutability. But in doing so BGH hold OP to a higher standard than Borjas (2003, 2014) holds himself. They also unmoor the regressions from their motivating theory, compromising interpretation.

Borjas (2003, section VII.A), introduces a nested constant-elasticity-of-substitution (CES) model with three levels:

$$Q_t = [\lambda_{Kt} K_t^\nu + \lambda_{Lt} L_t^\nu]^{1/\nu}$$

$$L_t = \left(\sum_i \theta_{it} L_{it}^\rho\right)^{1/\rho}$$

$$L_{it} = \left(\sum_j \alpha_{ijt} L_{ijt}^\eta\right)^{1/\eta}$$

where $t, i, j$ index time, education level, and experience level and in practice $\nu = 0$. Equating the cell-level wage $j$ to the marginal product of cell-level labor, we obtain

$$
\begin{aligned}
\log w_{ijt} &= \log \frac{\partial Q_t}{\partial L_{ijt}} \\
&= \log \lambda_{Lt} + \frac{1}{\sigma_{KL}} \log Q_t + \left(\frac{1}{\sigma_E} - \frac{1}{\sigma_{KL}}\right) \log L_t + \log \theta_{it} + \left(\frac{1}{\sigma_X} - \frac{1}{\sigma_E}\right) \log L_{it} + \log \alpha_{ijt} - \frac{1}{\sigma_X} \log L_{ijt}
\end{aligned}
\tag{1}
$$

where $\sigma_{KL} \equiv 1/(1-\nu)$, $\sigma_E \equiv 1/(1-\rho)$, $\sigma_X \equiv 1/(1-\eta)$ are the elasticities of substitution. Since there is a distinct and unknown $\alpha_{ijt}$ for every observation, they and the other parameters cannot all be identified. To achieve identification, Borjas (2003), p. 1361, assumes the $\alpha_{ijt}$ are time-invariant, fitting:

$$\log w_{ijt} = \delta_{it} + \delta_{ij} - \frac{1}{\sigma_X} \log L_{ijt} \tag{2}$$

where the $\delta_{it}$ capture the first five terms of (1) and the $\delta_{ij}$ model the $\log \alpha_{ijt}$. The fingerprint of the nested CES model in this empirical specification is the asymmetry in the choice of dummies. Out of conservatism, dummies $\delta_{jt}$ could be added to more fully model the $\alpha_{ijt}$ (or model omitted variables):

$$\log w_{ijt} = \delta_{it} + \delta_{ij} + \delta_{jt} - \frac{1}{\sigma_X} \log L_{ijt} \tag{3}$$

But this symmetrically dummy-saturated specification would effectively bring the empirical strategy back to the earlier, atheoretic part of Borjas (2003, eq. 3).

To appreciate this point, consider that Borjas (2003) could have swapped education and experience in his hierarchy. Subdividing labor first by experience then by education is certainly nonstandard, but all of these models are crude approximations, so none should be especially privileged. And OP find that doing this (in their "Model D") leads to very similar empirical results. By analogy with (2), the estimating equation for the bottom-level elasticity would then have been

Roodman, The domestic economic impacts of immigration

$$\log w_{ijt} = \delta_{jt} + \delta_{ij} - \frac{1}{\sigma_E}\log L_{ijt}.$$

If Borjas or BGH had then saturated this equation with dummies, it would have again brought us to (3)—but would now we would interpret the coefficient on $\log L_{ijt}$ as $-1/\sigma_E$ rather than $-1/\sigma_X$. Thus, *saturating hierarchical CES estimating equations with dummies eliminates their theoretical meaning*.

Arguably, the root of the trouble is that the hierarchical CES model is too general to be fit to the data, because labor's productivity, $\alpha_{ijt}$, may vary as much as its supply, $L_{ijt}$. The competition between the unobserved $\alpha_{ijt}$ and the observed $L_{ijt}$ to explain $w_{ijkt}$ forces an additional assumption, about how much variation to allow for $\alpha_{ijt}$. And that opens the door to dispute. Options that conservatively allow more variation in $\alpha_{ijt}$, such as including $\delta_{ij} + \delta_{jt}$ instead of just $\delta_{ij}$, remove parallel identifying variation in $L_{ijt}$. There is a bias-efficiency tradeoff.

Now, to the Borjas hierarchical model, OP add a fourth level, splitting natives from immigrants. Imitating Borjas (2003), the unidentified equation for the elasticity at this new bottom level, $\sigma_N$, would have the form:

$$\log w_{ijkt} = \delta_{ijt} + \log \alpha_{ijkt} - \frac{1}{\sigma_N}\log L_{ijkt} \tag{4}$$

where $k = F$ or $D$ for foreign or domestic. This would be estimated on a data set with observations by education, experience, immigration status, and census year. If OP had copied Borjas in assuming that the $\alpha_{ijkt}$ are time-invariant, they would have fit

$$\log w_{ijkt} = \delta_{ijt} + \delta_{ijk} - \frac{1}{\sigma_N}\log L_{ijkt} \tag{5}$$

However OP estimate differently. They subtract the $\log w_{ijDt}$ equation from the $\log w_{ijFt}$ one, both given by (4), to get

$$\log \frac{w_{ijFt}}{w_{ijDt}} = \frac{\log \alpha_{ijFt}}{\log \alpha_{ijDt}} - \frac{1}{\sigma_N}\log \frac{L_{ijFt}}{L_{ijDt}}$$

for which, in the Borjas (2003) mold, the estimating equation would be

$$\log \frac{w_{ijFt}}{w_{ijDt}} = \delta_{ij} - \frac{1}{\sigma_N}\log \frac{L_{ijFt}}{L_{ijDt}} \tag{6}$$

The $\delta$ dummies here lack an immigration status subscript $k$ because $\log \alpha_{ijFt}/\log \alpha_{ijDt}$ does not depend on $k$. They lack a $t$ subscript in analogy with Borjas (2003). As a result, estimating with this equation would be exactly as conservative in controlling for fixed effects as in Borjas (2003, 2014). In fact, it produces the same point estimates as (5).

But OP are more conservative because they add time dummies:

Roodman, The domestic economic impacts of immigration

$$\log \frac{w_{ijFt}}{w_{ijDt}} = \delta_{ij} + \delta_t - \frac{1}{\sigma_N} \log \frac{L_{ijFt}}{L_{ijDt}}$$

This allows $\log \alpha_{ijFt}/\log \alpha_{ijDt}$, the log ratio of the productivities for immigrant and native labor, to have two additive components, one varying over skill cells, one over time. This equation is used for columns 2, 3, 5, 6 of OP Table 2 and columns 3 and 7 of BGH Table 1.

Even though OP are more conservative than Borjas, BGH argue that OP are not conservative enough. More fixed-effect dummies are needed to rule out competing causal stories. Their robustness testing (columns 4 and 8 of BGH Table 1) adds 78 dummies to a regression with 240 observations:

$$\log \frac{w_{ijFt}}{w_{ijDt}} = \delta_{ij} + \delta_{it} + \delta_{jt} - \frac{1}{\sigma_N} \log \frac{L_{ijFt}}{L_{ijDt}} \tag{7}$$

But, as noted above, in recognizing and combating competing causal stories—in effect, in expanding their theory beyond the pure hierarchical CES model—BGH effectively detach their estimator from theory.

The recognition that BGH ask for more rigor in OP than is found in Borjas raises a question. What happens when the two sides' studies are made comparable? Do key Borjas results meet the BGH standard of rigor, with all those dummies? If OP cut back their dummy set to match Borjas, would their results strengthen?

Using the data and code for Borjas (2014)—the latest data and Borjas's preferred construction of variables and weights—I find that:

- Estimating $\sigma_N$, the elasticity of substitution between immigrants and natives, using the Borjas form (equation (5) or (6) above) strongly supports the OP finding that the elasticity is not infinite, but about 20. For just male workers, who are seen as a better proxy for earnings effects, $1/\sigma_N = 0.048$, standard error $= 0.01$, $p = 0.000$. For male and female workers together, $1/\sigma_N = 0.027$, standard error $= 0.011$, $p = 0.018$.[2]
- The BGH fixed effect–saturation robustness test based on (3) doubles the Borjas estimate of the elasticity of substitution across experience levels, indicating more diffuse and even wage impacts. The 95% confidence intervals now include infinity. Compared to Borjas (2014), Table 5.1, row 1, where $1/\sigma_X$ is put at 0.153 (standard error $= 0.060$) for men and 0.112 (standard error $= 0.051$) for men and women together, adding time-experience interaction dummies $\delta_{jt}$ reduces the estimates to 0.070 (standard error $= 0.049$) and 0.071 (standard error $= 0.050$). $F$ tests, to borrow a BGH argument, strongly favor keeping the new dummies.[3]
- As explained above, we can put Borjas's other parameter of interest, substitutability across education groups, to the same test if we invert education and experience in his model. Saturating with dummies results in exactly the same equation and results as just cited—now interpreted as being for $1/\sigma_E$.

In sum, under the Borjas (2003 and 2014) standard of rigor, OP appear correct about the partial complementary between immigrants and natives, even as they agree with Borjas (2014) on the other elasticities. The tougher BGH standard weakens the statistical strength and/or destroys the theoretical meaning of *all* the elasticity estimates in play in these studies. If that nihilism is warranted, then BGH seem mainly to undercut the skill cell literature initiated by Borjas (2003).

---

[2] The regression for men and women together corrects Borjas's mistaken use of just male wages for immigrants. Results are from regressions based on (5). Regressions based on (6) produce the same point estimates but smaller standard errors. The Stata command file accompanies this document.

[3] The Stata command file accompanies this document.